%
%
%
\def\unredoffs{} 

%
%
%
%
\newbox\leftpage \newdimen\fullhsize \newdimen\hstitle \newdimen\hsbody
\tolerance=1000\hfuzz=2pt
\catcode`\@=11 
\magnification=1200\unredoffs\baselineskip=16pt plus 2pt minus 1pt
\hsbody=\hsize \hstitle=\hsize 
%
%
%
%
%
\newcount\yearltd\yearltd=\year\advance\yearltd by -1900

%
%

\def\draftmode{\message{ DRAFTMODE }\def\draftdate{{\rm preliminary draft:
\number\month/\number\day/\number\yearltd\ \ \hourmin}}%
\headline={\hfil\draftdate}\writelabels\baselineskip=20pt plus 2pt minus 2pt
 {\count255=\time\divide\count255 by 60 \xdef\hourmin{\number\count255}
  \multiply\count255 by-60\advance\count255 by\time
  \xdef\hourmin{\hourmin:\ifnum\count255<10 0\fi\the\count255}}}
\def\nolabels{\def\wrlabeL##1{}\def\eqlabeL##1{}\def\reflabeL##1{}}
\def\writelabels{\def\wrlabeL##1{\leavevmode\vadjust{\rlap{\smash%
{\line{{\escapechar=` \hfill\rlap{\sevenrm\hskip.03in\string##1}}}}}}}%
\def\eqlabeL##1{{\escapechar-1\rlap{\sevenrm\hskip.05in\string##1}}}%
\def\reflabeL##1{\noexpand\llap{\noexpand\sevenrm\string\string\string##1}}}
\nolabels
%
\global\newcount\secno \global\secno=0
\global\newcount\meqno \global\meqno=1
\def\newsec#1{\global\advance\secno by1\message{(\the\secno. #1)}
\global\subsecno=0\eqnres@t\noindent{\bf\the\secno. #1}
\writetoca{{\secsym} {#1}}\par\nobreak\medskip\nobreak}
\def\eqnres@t{\xdef\secsym{\the\secno.}\global\meqno=1\bigbreak\bigskip}
\def\sequentialequations{\def\eqnres@t{\bigbreak}}\xdef\secsym{}
\global\newcount\subsecno \global\subsecno=0
\def\subsec#1{\global\advance\subsecno by1\message{(\secsym\the\subsecno. #1)}
\ifnum\lastpenalty>9000\else\bigbreak\fi
\noindent{\it\secsym\the\subsecno. #1}\writetoca{\string\quad 
{\secsym\the\subsecno.} {#1}}\par\nobreak\medskip\nobreak}
\def\appendix#1#2{\global\meqno=1\global\subsecno=0\xdef\secsym{\hbox{#1.}}
\bigbreak\bigskip\noindent{\bf Appendix #1. #2}\message{(#1. #2)}
\writetoca{Appendix {#1.} {#2}}\par\nobreak\medskip\nobreak}
%
%
\def\eqnn#1{\xdef #1{(\secsym\the\meqno)}\writedef{#1\leftbracket#1}%
\global\advance\meqno by1\wrlabeL#1}
\def\eqna#1{\xdef #1##1{\hbox{$(\secsym\the\meqno##1)$}}
\writedef{#1\numbersign1\leftbracket#1{\numbersign1}}%
\global\advance\meqno by1\wrlabeL{#1$\{\}$}}
\def\eqn#1#2{\xdef #1{(\secsym\the\meqno)}\writedef{#1\leftbracket#1}%
\global\advance\meqno by1$$#2\eqno#1\eqlabeL#1$$}
%
\newskip\footskip\footskip14pt plus 1pt minus 1pt 
\def\footnotefont{\ninepoint}\def\f@t#1{\footnotefont #1\@foot}
\def\f@@t{\baselineskip\footskip\bgroup\footnotefont\aftergroup\@foot\let\next}
\setbox\strutbox=\hbox{\vrule height9.5pt depth4.5pt width0pt}
\global\newcount\ftno \global\ftno=0
\def\foot{\global\advance\ftno by1\footnote{$^{\the\ftno}$}}
%
\newwrite\ftfile   
\def\footend{\def\foot{\global\advance\ftno by1\chardef\wfile=\ftfile
$^{\the\ftno}$\ifnum\ftno=1\immediate\openout\ftfile=foots.tmp\fi%
\immediate\write\ftfile{\noexpand\smallskip%
\noexpand\item{f\the\ftno:\ }\pctsign}\findarg}%
\def\footatend{\vfill\eject\immediate\closeout\ftfile{\parindent=20pt
\centerline{\bf Footnotes}\nobreak\bigskip\input foots.tmp }}}
\def\footatend{}
%
%
\global\newcount\refno \global\refno=1
\newwrite\rfile
\def\ref{[\the\refno]\nref}
\def\nref#1{\xdef#1{[\the\refno]}\writedef{#1\leftbracket#1}%
\ifnum\refno=1\immediate\openout\rfile=refs.tmp\fi
\global\advance\refno by1\chardef\wfile=\rfile\immediate
\write\rfile{\noexpand\item{#1\ }\reflabeL{#1\hskip.31in}\pctsign}\findarg}
\def\findarg#1#{\begingroup\obeylines\newlinechar=`\^^M\pass@rg}
{\obeylines\gdef\pass@rg#1{\writ@line\relax #1^^M\hbox{}^^M}%
\gdef\writ@line#1^^M{\expandafter\toks0\expandafter{\striprel@x #1}%
\edef\next{\the\toks0}\ifx\next\em@rk\let\next=\endgroup\else\ifx\next\empty%
\else\immediate\write\wfile{\the\toks0}\fi\let\next=\writ@line\fi\next\relax}}
\def\striprel@x#1{} \def\em@rk{\hbox{}} 
\def\lref{\begingroup\obeylines\lr@f}
\def\lr@f#1#2{\gdef#1{\ref#1{#2}}\endgroup\unskip}

\def\addref#1{\immediate\write\rfile{\noexpand\item{}#1}} 
\def\startrefs#1{\immediate\openout\rfile=refs.tmp\refno=#1}
\def\xref{\expandafter\xr@f}\def\xr@f[#1]{#1}
\def\refs#1{\count255=1[\r@fs #1{\hbox{}}]}
\def\r@fs#1{\ifx\und@fined#1\message{reflabel \string#1 is undefined.}%
\nref#1{need to supply reference \string#1.}\fi%
\vphantom{\hphantom{#1}}\edef\next{#1}\ifx\next\em@rk\def\next{}%
\else\ifx\next#1\ifodd\count255\relax\xref#1\count255=0\fi%
\else#1\count255=1\fi\let\next=\r@fs\fi\next}
%

%
\newwrite\ffile\global\newcount\figno \global\figno=1
\def\fig{fig.~\the\figno\nfig}
\def\nfig#1{\xdef#1{fig.~\the\figno}%
\writedef{#1\leftbracket fig.\noexpand~\the\figno}%
\ifnum\figno=1\immediate\openout\ffile=figs.tmp\fi\chardef\wfile=\ffile%
\immediate\write\ffile{\noexpand\medskip\noexpand\item{Fig.\ \the\figno. }
\reflabeL{#1\hskip.55in}\pctsign}\global\advance\figno by1\findarg}
\def\xfig{\expandafter\xf@g}\def\xf@g fig.\penalty\@M\ {}
\def\figs#1{figs.~\f@gs #1{\hbox{}}}
\def\f@gs#1{\edef\next{#1}\ifx\next\em@rk\def\next{}\else
\ifx\next#1\xfig #1\else#1\fi\let\next=\f@gs\fi\next}
\newwrite\lfile
{\escapechar-1\xdef\pctsign{\string\%}\xdef\leftbracket{\string\{}
\xdef\rightbracket{\string\}}\xdef\numbersign{\string\#}}

\def\writestop{\def\writestoppt{\immediate\write\lfile{\string\pageno%
\the\pageno\string\startrefs\leftbracket\the\refno\rightbracket%
\string\def\string\secsym\leftbracket\secsym\rightbracket%
\string\secno\the\secno\string\meqno\the\meqno}\immediate\closeout\lfile}}
\def\writestoppt{}\def\writedef#1{}
\def\seclab#1{\xdef #1{\the\secno}\writedef{#1\leftbracket#1}\wrlabeL{#1=#1}}
\def\subseclab#1{\xdef #1{\secsym\the\subsecno}%
\writedef{#1\leftbracket#1}\wrlabeL{#1=#1}}
\newwrite\tfile \def\writetoca#1{}
\def\leaderfill{\leaders\hbox to 1em{\hss.\hss}\hfill}
\def\writetoc{\immediate\openout\tfile=toc.tmp 
   \def\writetoca##1{{\edef\next{\write\tfile{\noindent ##1 
   \string\leaderfill {\noexpand\number\pageno} \par}}\next}}}

%
\catcode`\@=12 
%
\edef\tfontsize{\ifx\answ\bigans scaled\magstep3\else scaled\magstep4\fi}
\font\titlerm=cmr10 \tfontsize \font\titlerms=cmr7 \tfontsize
\font\titlermss=cmr5 \tfontsize \font\titlei=cmmi10 \tfontsize
\font\titleis=cmmi7 \tfontsize \font\titleiss=cmmi5 \tfontsize
\font\titlesy=cmsy10 \tfontsize \font\titlesys=cmsy7 \tfontsize
\font\titlesyss=cmsy5 \tfontsize \font\titleit=cmti10 \tfontsize
\skewchar\titlei='177 \skewchar\titleis='177 \skewchar\titleiss='177
\skewchar\titlesy='60 \skewchar\titlesys='60 \skewchar\titlesyss='60
\def\titlefont{\def\rm{\fam0\titlerm}
\textfont0=\titlerm \scriptfont0=\titlerms \scriptscriptfont0=\titlermss
\textfont1=\titlei \scriptfont1=\titleis \scriptscriptfont1=\titleiss
\textfont2=\titlesy \scriptfont2=\titlesys \scriptscriptfont2=\titlesyss
\textfont\itfam=\titleit \def\it{\fam\itfam\titleit}\rm}
 \ifx\answ\bigans\else scaled\magstep1\fi
\ifx\answ\bigans\else

 \font\absi=cmmi10 scaled\magstep1
\font\absis=cmmi7 scaled\magstep1 \font\absiss=cmmi5 scaled\magstep1
\font\abssy=cmsy10 scaled\magstep1 \font\abssys=cmsy7 scaled\magstep1
\font\abssyss=cmsy5 scaled\magstep1 
\skewchar\absi='177 \skewchar\absis='177 \skewchar\absiss='177
\skewchar\abssy='60 \skewchar\abssys='60 \skewchar\abssyss='60
\fi
\font\ninerm=cmr9 \font\sixrm=cmr6 \font\ninei=cmmi9 \font\sixi=cmmi6 
\font\ninesy=cmsy9 \font\sixsy=cmsy6 \font\ninebf=cmbx9 
\font\nineit=cmti9 \font\ninesl=cmsl9 \skewchar\ninei='177
\skewchar\sixi='177 \skewchar\ninesy='60 \skewchar\sixsy='60 
\def\ninepoint{\def\rm{\fam0\ninerm}
\textfont0=\ninerm \scriptfont0=\sixrm \scriptscriptfont0=\fiverm
\textfont1=\ninei \scriptfont1=\sixi \scriptscriptfont1=\fivei
\textfont2=\ninesy \scriptfont2=\sixsy \scriptscriptfont2=\fivesy
\textfont\itfam=\ninei \def\it{\fam\itfam\nineit}\def\sl{\fam\slfam\ninesl}%
\textfont\bffam=\ninebf \def\bf{\fam\bffam\ninebf}\rm} 
%
%
\def\noblackbox{\overfullrule=0pt}
\hyphenation{anom-aly anom-alies coun-ter-term coun-ter-terms}
\def\inv{^{\raise.15ex\hbox{${\scriptscriptstyle -}$}\kern-.05em 1}}

\def\Dsl{\,\raise.15ex\hbox{/}\mkern-13.5mu D} 
\def\dsl{\raise.15ex\hbox{/}\kern-.57em\partial}

 \def\Tr{{\rm Tr}}

\def\lspace{\ifx\answ\bigans{}\else\qquad\fi}
\def\lbspace{\ifx\answ\bigans{}\else\hskip-.2in\fi} 
\def\boxeqn#1{\vcenter{\vbox{\hrule\hbox{\vrule\kern3pt\vbox{\kern3pt
        \hbox{${\displaystyle #1}$}\kern3pt}\kern3pt\vrule}\hrule}}}
\def\mbox#1#2{\vcenter{\hrule \hbox{\vrule height#2in
                \kern#1in \vrule} \hrule}}  
%

\def\darr#1{\raise1.5ex\hbox{$\leftrightarrow$}\mkern-16.5mu #1}

\def\half{{\textstyle{1\over2}}} 
\def\roughly#1{\raise.3ex\hbox{$#1$\kern-.75em\lower1ex\hbox{$\sim$}}}

\openup -1pt
\expandafter\ifx\csname pre amssym.tex at\endcsname\relax \else\endinput\fi
\expandafter\chardef\csname pre amssym.tex at\endcsname=\the\catcode`\@
\catcode`\@=11
\ifx\undefined\newsymbol \else \begingroup\def\input#1 {\endgroup}\fi
\expandafter\ifx\csname amssym.def\endcsname\relax \else\endinput\fi
\expandafter\edef\csname amssym.def\endcsname{%
       \catcode`\noexpand\@=\the\catcode`\@\space}
\catcode`\@=11
\def\undefine#1{\let#1\undefined}
\def\newsymbol#1#2#3#4#5{\let\next@\relax
 \ifnum#2=\@ne\let\next@\msafam@\else
 \ifnum#2=\tw@\let\next@\msbfam@\fi\fi
 \mathchardef#1="#3\next@#4#5}
\def\mathhexbox@#1#2#3{\relax
 \ifmmode\mathpalette{}{\m@th\mathchar"#1#2#3}%
 \else\leavevmode\hbox{$\m@th\mathchar"#1#2#3$}\fi}
\def\hexnumber@#1{\ifcase#1 0\or 1\or 2\or 3\or 4\or 5\or 6\or 7\or 8\or
 9\or A\or B\or C\or D\or E\or F\fi}
\font\tenmsa=msam10
\font\sevenmsa=msam7
\font\fivemsa=msam5
\newfam\msafam
\textfont\msafam=\tenmsa
\scriptfont\msafam=\sevenmsa
\scriptscriptfont\msafam=\fivemsa
\edef\msafam@{\hexnumber@\msafam}
\mathchardef\dabar@"0\msafam@39
\def\maltese{{\mathhexbox@\msafam@7A}}
\font\tenmsb=msbm10
\font\sevenmsb=msbm7
\font\fivemsb=msbm5
\newfam\msbfam
\textfont\msbfam=\tenmsb
\scriptfont\msbfam=\sevenmsb
\scriptscriptfont\msbfam=\fivemsb
\edef\msbfam@{\hexnumber@\msbfam}
\def\Bbb#1{{\fam\msbfam\relax#1}}
\def\widehat#1{\setbox\z@\hbox{$\m@th#1$}%
 \ifdim\wd\z@>\tw@ em\mathaccent"0\msbfam@5B{#1}%
 \else\mathaccent"0362{#1}\fi}
\def\widetilde#1{\setbox\z@\hbox{$\m@th#1$}%
 \ifdim\wd\z@>\tw@ em\mathaccent"0\msbfam@5D{#1}%
 \else\mathaccent"0365{#1}\fi}
\font\teneufm=eufm10
\font\seveneufm=eufm7
\font\fiveeufm=eufm5
\newfam\eufmfam
\textfont\eufmfam=\teneufm
\scriptfont\eufmfam=\seveneufm
\scriptscriptfont\eufmfam=\fiveeufm

\csname amssym.def\endcsname
\relax
\newsymbol\smallsetminus 2272
\noblackbox
\newcount\figno
\figno=0
\def\mathrm#1{{\rm #1}}
\def\fig#1#2#3{
\par\begingroup\parindent=0pt\leftskip=1cm\rightskip=1cm\parindent=0pt
\baselineskip=11pt
\global\advance\figno by 1
\midinsert
\epsfxsize=#3
\centerline{\epsfbox{#2}}
\vskip 12pt
\centerline{{\bf Figure \the\figno} #1}\par
\endinsert\endgroup\par}
\font\tenmsb=msbm10       \font\sevenmsb=msbm7
\font\fivemsb=msbm5       \newfam\msbfam
\textfont\msbfam=\tenmsb  \scriptfont\msbfam=\sevenmsb
\scriptscriptfont\msbfam=\fivemsb
\def\Bbb#1{{\fam\msbfam\relax#1}}

\def\Zop{{\Bbb Z}}

\def\bbbc{{\mathchoice {\setbox0=\hbox{$\displaystyle\rm C$}\hbox{\hbox
to0pt{\kern0.4\wd0\vrule height0.9\ht0\hss}\box0}}
{\setbox0=\hbox{$\textstyle\rm C$}\hbox{\hbox
to0pt{\kern0.4\wd0\vrule height0.9\ht0\hss}\box0}}
{\setbox0=\hbox{$\scriptstyle\rm C$}\hbox{\hbox
to0pt{\kern0.4\wd0\vrule height0.9\ht0\hss}\box0}}
{\setbox0=\hbox{$\scriptscriptstyle\rm C$}\hbox{\hbox
to0pt{\kern0.4\wd0\vrule height0.9\ht0\hss}\box0}}}}
\def\figlabel#1{\xdef#1{\the\figno}}

\def\pmb#1{\setbox0=\hbox{#1}%
 \kern-.025em\copy0\kern-\wd0
 \kern.05em\copy0\kern-\wd0
 \kern-.025em\raise.0433em\box0 }

\def\half{{1\over 2}}



\def\H{{\cal H}}

\def\A{{\cal{A}}}

\def\Tr{{\hbox{Tr}}}

\def\smallmatrix{\raise1.6pt\hbox{${\scriptscriptstyle 
{{\;0\, \; 1} \choose {-\!1\;0}} 
}$}}
\def\smallmatrixone{\raise1.6pt\hbox{${\scriptscriptstyle 
{{1\,\;0} \choose {\,0\; \, 1}} 
}$}}
\def\hsmallsetminus{\hbox{\raise1.5pt\hbox{$\smallsetminus$}}}
\def\tilM{\hbox{${\scriptstyle \widetilde{\phantom M}}$}\hskip-9pt
   \raise1.3pt\hbox{${\scriptstyle M}$}\,}
\def\sn{\smallskip\noindent}
\def\mn{\medskip\noindent}
\def\bn{\bigskip\noindent}
\def\hbn{\hfill\break\noindent}


\def\figin{\epsfcheck\figin}\def\figins{\epsfcheck\figins}
\def\epsfcheck{\ifx\epsfbox\UnDeFiNeD
\message{(NO epsf.tex, FIGURES WILL BE IGNORED)}
\gdef\figin##1{\vskip2in}\gdef\figins##1{\hskip.5in}
\else\message{(FIGURES WILL BE INCLUDED)}%
\gdef\figin##1{##1}\gdef\figins##1{##1}\fi}
\def\DefWarn#1{}
\def\figinsert{\goodbreak\midinsert}
\def\ifig#1#2#3{\DefWarn#1\xdef#1{fig.~\the\figno}
\writedef{#1\leftbracket fig.\noexpand~\the\figno}%
\figinsert\figin{\centerline{#3}}\medskip\centerline{\vbox{\baselineskip12pt
\advance\hsize by -1truein\noindent\footnotefont{\bf Fig.~\the\figno:} #2}}
\bigskip\endinsert\global\advance\figno by1}


\def\summand{\sum_{(J,\nu \kappa)\atop{\rm{s.t.}\,2J+\nu \kappa\,\rm{even}}}}
\def\summandone{\sum_{(J',n')\atop{\rm{s.t.}\,2J'+n'}\,\rm{even}}}

\def\bracket{(J,\nu \kappa)}
\def\bracketone{({k\over 2}-J,\nkt+k)}

\def\kt{\kappa}
\def\nkt{\nu \kappa}
\def\n{\nu }
\def\k{\kappa}

\def\virquantity{ \overline{D}{}^{j}_{r,s}(g_\alpha)\,D^{j}_{r,s}(g_\beta)\,\chi^{\rm{Vir}}_{j^2}(\tilde{q})}

\def\a{\alpha}
\def\b{\beta}
\def\summandthree{\sum_{\n=-\k+1}^{\k}}
\def\virsum{\sum_{{{{{j \in{1\over 2}\Zop_+}\atop{r+s=\k\rho+\n}}}\atop{r-s=\k\rho'}}\atop{\rm{s.t. \rho+\rho'\rm{even}}}}}
\def\ssu{S^{{\rm SU(2)}_k}}
\def\bssu{\bar{S}^{{\rm SU(2)}_k}}

\def\suk{{\scriptscriptstyle {\rm SU(2)}_k}}

\magnification=\magstep1 \overfullrule=0pt
\advance\hoffset by -0.3truecm  
\def\frac#1#2{{#1\over#2}}

\def\rra{\rangle\!\rangle}


\lref\GR{
  M.R.~Gaberdiel, A.~Recknagel,
  {\it Conformal boundary states for free bosons and fermions},
  J.\ High Energy Phys.\  {\bf 0111}, 016 (2001); {\tt hep-th/0108238}.}

\lref\RSmod{A.\ Recknagel, V.\ Schomerus, {\it Boundary deformation
theory and moduli spaces of D-branes}, Nucl.\ Phys.\ {\bf B545}, 233
(1999); {\tt hep-th/9811237}.}

\lref\klebpol{I.R.\ Klebanov, A.M.\ Polyakov, {\it Interaction of
discrete states in two-dimensional string theory},
Mod.\ Phys.\ Lett.\ {\bf A6}, 3273 (1991); {\tt hep-th/9109032}.}

\lref\cklm{C.G.\ Callan, I.R.\ Klebanov, A.W.\ Ludwig, J.M.\
Maldacena, {\it Exact solution of a boundary conformal field theory},
Nucl.\ Phys.\ {\bf B422}, 417 (1994); {\tt hep-th/9402113}.}

\lref\polthor{J.\ Polchinski, L.\ Thorlacius, {\it Free fermion
representation of a boundary conformal field theory}, Phys.\ Rev.\
{\bf D50}, 622 (1994); {\tt hep-th/9404008}.}

\lref\fk{I.B.\ Frenkel, V.G.\ Kac, {\it Basic representations of
affine Lie algebras and dual resonance models}, Invent.\ Math.\
{\bf 62}, 23 (1981).}

\lref\segal{G.B.\ Segal, {\it Unitary representations of some infinite
dimensional groups}, Commun.\ Math.\ Phys.\ {\bf 80}, 301 (1981).}

\lref\Ham{M.\ Hamermesh, {\it Group theory and its applications to
physical problems}, Addison-Wesley (1962).}

\lref\cardy{J.L.\ Cardy, {\it Boundary conditions, fusion rules and
the Verlinde formula}, Nucl.\ Phys.\ B {\bf324}, 581 (1989).}

\lref\DGH{L.\ Dixon, P.\ Ginsparg, J.\ Harvey, {\it $\hat c=1$
superconformal field theory}, Nucl.\ Phys.\ {\bf B306}, 470 (1988).}

\lref\bg{O.\ Bergman, M.R.\ Gaberdiel, {\it A non-supersymmetric open
string theory and S-duality}, Nucl.\ Phys.\ {\bf B499}, 183 (1997);
{\tt hep-th/9701137}.}

\lref\feiginfuchs{B.L.\ Feigin, D.B.\ Fuchs, {\it Invariant
skew-symmetric differential operators on the line and Verma modules
over the Virasoro algebra}, Funct. Anal. Appl. {\bf 16}, 114 (1982);
\quad {\it Verma modules over the Virasoro algebra}, in:\ Lecture
Notes in Mathematics {\bf 1060}, Springer 1984.}

\lref\GRW{M.R.\ Gaberdiel, A.\ Recknagel, G.M.T.\ Watts,
{\it The conformal boundary states for SU(2) at level $1$};
{\tt hep-th/0108102}.}

\lref\knapp{A.W.\ Knapp, {\it Representation theory of semisimple
groups: an overview based on examples}, Princeton University Press,
Princeton (1986).}

\lref\cohnfried{J.D.\ Cohn, D.\ Friedan, {\it Super characters and
chiral asymmetry in superconformal field theory}, Nucl.\ Phys.\
{\bf B296}, 779 (1988).}

\lref\DixHar{L.J.\ Dixon, J.A.\ Harvey, {\it String theories
in ten dimensions without spacetime supersymmetry}, Nucl.\ Phys.\
{\bf B274}, 93 (1986).}

\lref\seiwitt{N.\ Seiberg, E.\ Witten, {\it Spin structures
in string theory}, Nucl.\ Phys.\ {\bf B276}, 272 (1986).}

\lref\sen{A.\ Sen, {\it BPS D-branes on non-supersymmetric
cycles}, J.\ High Energy Phys.\  {\bf 9812} (1998) 021;
{\tt hep-th/9812031}.}

\lref\arfken{G.\ Arfken, {\it Mathematical methods for physicists},
Academic Press (1970).}

\lref\Friedan{D.\ Friedan, {\it The space of conformal boundary
conditions for the $c=1$ Gaussian model}, unpublished note (1999).}

\lref\gutperle{M.\ Gutperle, {\it Non-BPS D-branes and enhanced
symmetry in an asymmetric orbifold}, J.\ High Energy Phys.\ {\bf 0008}
(2000) 036; {\tt hep-th/0007126}.}

\lref\Janik{R.A.\ Janik, {\it Exceptional boundary states at $c=1$},
  Nucl.\ Phys.\  B {\bf 618} (2001) 675; {\tt hep-th/0109021}.
}

\lref\thompson{D.M.\ Thompson, {\it Descent relations in type 0A/0B},
{\tt hep-th/0105314}.}

\lref\MMS{J.\ Maldacena, G.\ Moore, N.\ Seiberg, {\it  Geometrical
interpretation of D-branes in gauged WZW models}, J.\ High Energy Phys.\  {\bf 0107} 
(2001) 046; {\tt hep-th/0105038}.}

\lref\Sark{
  G.\ Sarkissian, 
  {\it Non-maximally symmetric D-branes on group manifold in 
    the Lagrangian approach}, J.\ High Energy Phys.\ 
      {\bf 0207} (2001) 033 
   {\tt hep-th/0205097}. 
}

\lref\MMStwo{
  J.M.~Maldacena, G.W.~Moore, N.~Seiberg,
  {\it D-brane instantons and K-theory charges},
  J.\ High Energy Phys.\  {\bf 0111} (2001) 062; {\tt hep-th/0108100}.
}

\lref\fsb{L.\ Birke, J.\ Fuchs, C.\ Schweigert, {\it Symmetry breaking
boundary conditions and WZW orbifolds}, Adv.\ Theor.\ Math.\ Phys.\
{\bf 3}, 671 (1999); {\tt hep-th/9905038}.}

\lref\pol{J.\ Polchinski, {\it Combinatorics of boundaries in string
theory}, Phys.\ Rev.\ {\bf D50}, 6041 (1994); {\tt hep-th/9407031}.}

\lref\PolD{
J.\ Polchinski, {\it Dirichlet branes and Ramond-Ramond
 charges}, Phys.\ Rev.\ Lett.\ {\bf 75} (1995) 4724: 
 {\tt  hep-th/9510017}.
}

\lref\FS{J.\ Fuchs, C.\ Schweigert, {\sl A classifying algebra for
boundary conditions}, Phys.\ Lett.\ {\bf B414}, 251 (1997);
{\tt hep-th/9708141}.}

\lref\PSS{G.\ Pradisi, A.\ Sagnotti, Y.S.\ Stanev, {\it Completeness
conditions for boundary operators in 2d conformal field theory},
Phys.\ Lett.\ {\bf B381}, 97 (1996); {\tt hep-th/9603097}.}

\lref\BPPZ{R.E.\ Behrend, P.A.\ Pearce, V.B.\ Petkova, J.-B.\ Zuber,
{\it Boundary conditions in rational conformal field theories},
Nucl.\ Phys.\ B {\bf 579}, 707 (2000); {\tt hep-th/9908036}.}

\lref\Lew{D.C.\ Lewellen, {\it Sewing constraints for conformal field
theories on surfaces with boundaries}, Nucl.\ Phys.\ {\bf B372}, 654
(1992).}

\lref\GG{M.B.\ Green, M.\ Gutperle, {\it Symmetry breaking at
enhanced symmetry points}, Nucl.\ Phys.\ {\bf B460}, 77 (1996);
{\tt hep-th/9509171}.}

\lref\QS{
 T.~Quella, V.~Schomerus,
  {\it Symmetry breaking boundary states and defect lines},
  J.\ High Energy Phys.\  {\bf 0206} (2002) 028; {\tt hep-th/0203161}.
}

\lref\FuchsLect{
 J.~Fuchs,
 {\it Lectures on conformal field theory and Kac-Moody algebras};
 {\tt hep-th/9702194}.
}

\lref\FredGab{
  S.~Fredenhagen, M.~R.~Gaberdiel, C.A.~Keller,
  {\it Bulk induced boundary perturbations},
  J.\ Phys.\ A  {\bf 40}, F17 (2007); 
  {\tt hep-th/0609034}.
}
\lref\HemmingDM{
  S.~Hemming, S.~Kawai, E.~Keski-Vakkuri,
  {\it Coulomb-gas formulation of SU(2) branes and chiral blocks},
  J.\ Phys.\ A  {\bf 38}, 5809 (2005); 
  {\tt hep-th/0403145}.
}

\lref\DistlerXV{
  J.~Distler, Z.A.~Qiu,
  {\it BRS cohomology and a Feigin-Fuchs representation of Kac-Moody and
  parafermionic theories},
  Nucl.\ Phys.\  B {\bf 336}, 533 (1990).
}

\lref\Ishi{
   N.\ Ishibashi, {\it The boundary and crosscap states
   in conformal field theories}, Mod.\ Phys.\ Lett.\ A{\bf4} (1989) 251.
}

\lref\Behthree{
  R.E.\ Behrend, P.A.\ Pearce, V.B.\ Petkova, J.-B.\ Zuber,  {\it
  Boundary conditions in \hbox{rational} conformal field theories},
  Nucl.\ Phys.\ B {\bf570} (2000) 525, {\bf579} (2000) 707; 
  {\tt hep-th/9908036}.
}

\lref\PetkovaZuber{
 V.B.~Petkova, J.B.~Zuber,
  {\it Conformal boundary conditions and what they teach us},
 {\tt hep-th/0103007}.
}

\lref\QuellaNS{
  T.~Quella,
  {\it On the hierarchy of symmetry breaking D-branes in group manifolds},
  J.\ High Energy Phys.\  {\bf 0212}, 009 (2002); {\tt hep-th/0209157}.
}

\lref\QuellaFK{
  T.~Quella, V.~Schomerus,
  {\it Asymmetric cosets},
  J.\ High Energy Phys.\  {\bf 0302} (2003) 030; {\tt hep-th/0212119}.
}

\lref\Monnier{ 
S.~Monnier,
  {\it D-branes in Lie groups of rank $>$ 1},
  J.\ High Energy Phys.\  {\bf 0508} (2005) 062; {\tt hep-th/0507159}.
}

\lref\ARS{
 A.Yu.\ Alekseev, A.\ Recknagel, V.\ Schomerus,
  {\it Brane dynamics in background fluxes and non-commuta\-tive
  geometry}, J.\ High Energy Phys.\ 05 (2000) 010; {\tt hep-th/0003187}.
}

\lref\ForsteKM{
  S.~F\"orste, D.~Roggenkamp,
  {\it Current current deformations of conformal field theories, and WZW
  models}, J.\ High Energy Phys.\  {\bf 0305} (2003) 071; {\tt hep-th/0304234}.
}

\lref\HassanGI{
  S.F.~Hassan, A.~Sen,
  {\it Marginal deformations of WZNW and coset models from O(D,D)
  transformation}, Nucl.\ Phys.\  B {\bf 405} (1993) 143; {\tt 
  hep-th/9210121}.
}

\lref\CHS{C.G.~Callan, J.~A.~Harvey, A.~Strominger,
  {\it World sheet approach to heterotic instantons and solitons},
  Nucl.\ Phys.\  B {\bf 359} (1991) 611; \ \ 
  {\it Supersymmetric string solitons}, {\tt hep-th/9112030}.
}

\lref\BianchiGT{
  M.~Bianchi, Y.S.~Stanev,
  {\it Open strings on the Neveu-Schwarz pentabrane},
  Nucl.\ Phys.\  B {\bf 523} (1998) 193; {\tt hep-th/9711069}.
}

\lref\FFFS{
G.~Felder, J.~Fr\"ohlich, J.~Fuchs, C.~Schweigert,
  {\it The geometry of WZW branes},
  J.\ Geom.\ Phys.\  {\bf 34} (2000) 162; {\tt hep-th/9909030}.
}

\lref\AlSchom{
A.Yu.~Alekseev, V.~Schomerus,
  {\it D-branes in the WZW model}, Phys.\ Rev.\  D {\bf 60} (1999) 061901; 
  {\tt hep-th/9812193}.
}

\lref\ReSchGep{
A.~Recknagel, V.~Schomerus,
  {\it D-branes in Gepner models},
  Nucl.\ Phys.\  B {\bf 531} (1998) 185; {\tt hep-th/9712186}.
}

\lref\BruDoug{
I.~Brunner, M.R.~Douglas, A.E.~Lawrence, C.~Romelsberger,
  {\it D-branes on the quintic},
  J.\ High Energy Phys.\  {\bf 0008} (2000) 015; 
  {\tt hep-th/9906200}.
}

\lref\Sen{
A.\ Sen, {\sl SO(32) spinors of type I and other solitons on
   brane-antibrane pair}, J.\ High Energy Phys.\ {\bf9809} (1998) 023; 
  {\tt hep-th/9808141}
}

\lref\AffL{
  I.\  Affleck, A.W.W.\  Ludwig, 
   {\it Critical theory of overscreened Kondo fixed points},
      Nucl.\ Phys.\ B {\bf360} (1991) 641; 
      \quad 
 {\sl The Kondo effect, conformal field theory and fusion rules},
   Nucl.\ Phys.\ B {\bf352} (1991) 849
}

\lref\FriKon{
D.~Friedan, A.~Konechny,
  {\it On the boundary entropy of one-dimensional quantum systems at low
  temperature},   Phys.\ Rev.\ Lett.\  {\bf 93} (2004) 030402; (2004)
  {\tt hep-th/0312197}.
}

\lref\Fredenhagen{
S.~Fredenhagen,
  {\it Organizing boundary RG flows},   Nucl.\ Phys.\  B {\bf 660} (2003) 
   436; {\tt hep-th/0301229}.
}

\lref\FredSchom{
S.~Fredenhagen, V.~Schomerus,
  {\it Branes on group manifolds, gluon condensates, and twisted K-theory},
  J.\ High Energy Phys.\ {\bf 0104} (2001) 007; {\tt hep-th/0012164}.
}

\lref\Runkel{
I.~Runkel,
  {\it Boundary structure constants for the A-series Virasoro minimal models},
  Nucl.\ Phys.\  B {\bf 549} (1999) 563;   {\tt hep-th/9811178}.
  }
  
\lref\FZZ{
  V.~Fateev, A.B.~Zamolodchikov, Al.B.~Zamolodchikov,
  {\it Boundary Liouville field theory. I: Boundary state and boundary  two-point
  function},
  {\tt hep-th/0001012}.
  }  

\lref\PonsTesch{
B.~Ponsot, J.~Teschner,
  {\it Boundary Liouville field theory: Boundary three point function},
  Nucl.\ Phys.\  B {\bf 622} (2002) 309; {\tt hep-th/0110244}.
}  

\lref\JoergRevisited{
J.~Teschner, {\it Liouville theory revisited},
  Class.\ Quant.\ Grav.\  {\bf 18} (2001) R153; {\tt hep-th/0104158}.
}  

\lref\ZamolodchikovAH{
 A.B.~Zamolodchikov, Al.B.~Zamolodchikov,
  {\it Liouville field theory on a pseudosphere}, {\tt hep-th/0101152}.
}

\lref\JoergRemarks{
  J.~Teschner,
  {\it Remarks on Liouville theory with boundary}, {\tt hep-th/0009138}.
}  

\def\tskh{{\textstyle {k\over2}}}

{\nopagenumbers
\line{arXiv:0705.1068 [hep-th] \hfill KCL-MTH-07-03}
\vskip3cm
\centerline{{\titlefont Symmetry-breaking boundary states}}
\vskip.5cm
\centerline{{\titlefont for WZW models}}
\vskip1.3cm
\centerline{{\bf Daniel Blakeley}\quad and\quad {\bf Andreas Recknagel}}
\vskip.7cm
\centerline{King's College London, Department of Mathematics}
\smallskip
\centerline{Strand, London WC2R 2LS, UK}
\smallskip
\centerline{\tt daniel.blakeley@kcl.ac.uk, andreas.recknagel@kcl.ac.uk}
\vskip1.8cm
\centerline{\bf Abstract}
\bigskip

{\narrower\narrower \noindent
Starting with the SU(2)$_k$ WZW model, we construct boundary states that generically preserve only a parafermion times Virasoro subalgebra of 
the full affine Lie algebra symmetry of the bulk model. The boundary 
states come in families: intervals for generic $k$, quotients of SU(2) by 
discrete groups if $k$ is a square. In that case, special members of the 
families can be viewed as superpositions of rotated Cardy branes. 
Using embeddings of SU(2) into higher groups, the new boundary 
states can be lifted to symmetry-breaking branes for other WZW models. 
\eject}

}

\pageno=1

\parindent=0pt

\newsec{Introduction}

D-branes have become an extremely important ingredient of 
string theory. Since their discovery \PolD, it has been 
clear that they have a world-sheet description as conformal 
boundary conditions, or boundary states. The CFT approach 
is distinguished from the target space picture of branes 
in that it does not refer to classical geometry -- which 
makes it harder to interpret its results, but broadens the 
scope towards D-branes that would be hard to find based 
on classical intuitions. 

The world-sheet construction and classification of D-branes 
is rather well under control if one restricts to boundary 
conditions that preserve the maximal symmetry, in rational 
conformal field theories \refs{\cardy,\Runkel}. If the CFT in question 
has a sigma model interpretation, one can often relate 
the CFT boundary states to D-brane submanifolds in the 
target; in WZW models, e.g., maximally symmetric boundary 
states correspond to conjugacy classes (perhaps rotated or 
twisted) in the group target \AlSchom. For other rational 
backgrounds like Gepner models \ReSchGep, the relation may 
already be more intricate due to lines of marginal stability 
in the bulk moduli space \BruDoug. 
\hbn
Consistency of string models on world-sheets with boundary 
only requires branes to preserve conformal symmetry, and 
therefore it is natural to study symmetry-breaking boundary 
states. However, there are at present no general principles 
to make up for the loss of symmetry, thus it is very 
difficult to construct symmetry-breaking boundary conditions. 
Apart from Virasoro minimal models, complete lists of conformal 
boundary states are known only for $c=1$ theories 
\refs{\Friedan,\GRW,\GR,\Janik} -- see also the earlier works
\refs{\cklm,\polthor,\RSmod} -- and (most probably) for the 
Liouville theory \refs{\FZZ,\JoergRemarks,\ZamolodchikovAH,
\JoergRevisited,\PonsTesch} 
(where it is more difficult to 
decide completeness due to the non-compactness of the model).
\hbn
Partially symmetry-breaking boundary state for WZW and 
coset models were studied in particular in \refs{\MMS,\QS}.
The maximal symmetry algebra is broken up into a subalgebra 
and its commutant, so that one can in particular choose
twisted gluing conditions for the subalgebra (and standard 
gluing conditions for the commutant).
\hbn
Finding a geometric interpretation of symmetry-breaking branes 
is usually difficult, though \MMS\ provides a target picture 
for special cases; see also \Sark. On the other hand, 
one expects boundary conditions with reduced symmetry 
to appear naturally in connection with boundary renormalisation 
group flows, such as tachyon condensations. For example, 
it turns out that Sen's process of 
dimensional transmutation \Sen\ can be described using 
conformal free boson boundary states \refs{\RSmod,\GR}. 
We hope that the families of symmetry-breaking SU(2) 
boundary states constructed in this article will find 
similar applications to string theory (e.g.\ in relation 
to branes in the NS 5-brane background \CHS) or in 
condensed matter physics (where SU(2) boundary states 
were used to solve the Kondo effect \AffL). 
\hbn
Apart from abstract results like the $g$-theorem \FriKon, too 
little is known at present about the details of RG flows to 
turn their study into an efficient method to construct 
symmetry-breaking boundary states.  Instead, we will exploit 
and extend the methods from  \refs{\MMS,\QS} here. 
\mn
The paper is organised as follows: As the construction of 
our boundary states rests on a decomposition 
of the SU(2)$_k$ state space into parafermions and free bosons, 
we start by reviewing this in some detail. In Section 3, we 
write down an ansatz for symmetry-breaking boundary states 
and show that it satisfies Cardy's condition, i.e.\ that 
the overlaps of two such boundary states can be regarded as 
an open string partition function. The analysis is performed 
for the case that the level $k$ of the WZW model is a square, 
while Section 4 deals with the general case, where a smaller 
family of boundary states results. In Section 5, we show 
that, for special values of the parameters, the new boundary 
states can be viewed as intersecting configurations of 
maximally symmetric SU(2)$_k$ boundary states; it appears
that the new families of boundary states interpolate between 
branes of different dimension in target space. Using 
embeddings of SU(2), we generate symmetry-breaking boundary 
states for higher rank WZW targets $G$ in Section 6, before 
concluding with a list of open problems.

\newsec{Decomposition of representations and Ishibashi states}

The idea of our construction is very simple: start from the bulk 
Hilbert space of the SU(2) WZW model at  level $k$ (with diagonal 
modular invariant  partition function), decompose each SU(2)$_k$ 
irrep into (sums of) products of parafermion times U(1) irreps, 
then decompose the latter further into Virasoro irreps -- both in 
the left- and the right-moving sector. Suitable left-right combinations
then provide Ishibashi states preserving the reduced parafermion 
times Virasoro symmetry. In the next section, we will propose 
symmetry-breaking boundary states as linear combinations of 
those Ishibashi states and show that Cardy's conditions are 
satisfied. Here, we review the decompositions of representations
which provide the Ishibashi states. 

\medskip
We work with an SU(2) WZW model with diagonal modular 
invariant bulk partition function; the bulk state space is 
$$
H = 
\bigoplus_{J=0,\ldots,{k\over 2}}H^{{\rm SU(2)}_k}_J\otimes \bar{H}^{{\rm SU(2)}_k}_J
\ \ .
$$
The irreducible SU(2)$_k$ representations $H^{{\rm SU(2)}_k}_J$ can 
be decomposed with respect to the smaller parafermion times free 
boson symmetry algebra ${\rm SU(2)}_k/{\rm U(1)}_k \ \otimes U(1)_k$, 
where more precisely U(1)$_k$ denotes the abelian current algebra 
at radius $r=\sqrt{k}\,r_{s.d.}$, extended by the local fields 
$\exp(\pm i \sqrt{2k}\, X(z))$ -- using the conventions of 
\MMS. 
The chiral algebra $\A$ of the parafermion theory has  irreducible representations 
labelled by $(J,n)$, where $J\in{1\over 2}\Zop$ with $0\leq J\leq {k\over 2}$  
and where $n=-k+1, \ldots, k$ is integer such that $2J+n$ is even; there is 
a field identification 
\eqn\fieldid{(\tskh-J, n+k)\sim (J,n) \ .}

The chiral SU(2)$_k$ modules decompose as 
$$
H^{{\rm SU(2)}_k}_J = \bigoplus_{{n \atop 2J+n \;{\rm even}}} 
  H^{{\rm PF}}_{(J,n)}  \otimes H^{{\rm U(1)}_k}_n
$$
which implies, at the level of characters, that
\eqn\character{
\chi_{J}^{{\rm SU(2)}_k}(q,z) =
\sum_{n=-k+1,\ldots, k\atop{2J+n\;{\rm even}}}
 \chi^{\rm PF}_{(J,n)}(q) \; \chi^{{\rm U(1)}_k}_{n}(q,z)\ \ .
 }
We will not need the explicit form of the parafermion characters $\chi^{PF}_{(J,n)}(q)$, which is for example given \refs{\DistlerXV,\HemmingDM,\MMS}, but the symmetries
\eqn\pfidentity{\chi^{PF}_{(J,n)}(q) \equiv \chi^{PF}_{({k\over 2}-J, n+k)}(q) = 
\chi^{PF}_{(J,-n)}(q)}
will be important later on. 

 The $U(1)_k$ characters are given by
\eqn\uonek{\chi_{n}^{U(1)_k}(q,z)={\Theta_{n,k}(q,z)\over \eta(q)}={1\over{\eta(q)}}\sum_{m\in\Zop}q^{k(m+{n\over{{2k}}})^2}
e^{2\pi izk(l+{n\over 2k})}}
where $\eta(q) = q^{1\over 24}\prod_{n=1}^{\infty}(1-q^n)$ is the Dedekind 
eta function. 
In particular, we have a decomposition 
\eqn\uone{
\chi_{n}^{U(1)_k}(q)= {1\over{\eta(q)}}\sum_{m\in\Zop}q^{k(m+{n\over{{2k}}})^2}
= \sum_{m\in\Zop} \chi_{m,n}^{U(1)}(q)
}
into characters of irreducible U(1) representations  $H^{{\rm U(1)}}_{m,n}$ built up over ground states of conformal dimension 
\eqn\confdim{h_{m,n}=k\Bigl(m+{n\over{2k}}\Bigr)^2\ \ .}

\medskip

To be able to construct boundary states which break the symmetry down to 
${\rm PF}\times {\rm Vir}$, we need to further decompose the U(1) 
representations into Virasoro representations:
At central charge $c=1$, every Virasoro Verma module is irreducible 
except if the highest weight is the square of a half-integer; therefore only 
irreducible U(1)-representation with highest weight $h_{(m,n)}=j^2$ for some 
$j\in\half\Zop$ are reducible with respect to the Virasoro algebra. 
Comparing to \confdim, we see that for generic $k$ 
this can only occur in the vacuum sector $m=n=0$; alternatively  we  
need that the level $k$ of the SU(2)$_k$ model is a square 
\eqn\kcondition{
k = \kappa^2\quad {\rm for\ some}\ \ \kappa \in \Zop_+
} 
and that the U(1)$_k$ representation label is 
\eqn\ncondition{
n = \nu  \cdot \kappa \quad {\rm for\ some}\ \ 
 \nu  = -\kappa +1, \ldots, \kappa\  .
}
We will address the case of generic $k$ in Section 4 below, but 
for the time being we assume that the two conditions 
\kcondition\ and \ncondition\ above are satisfied. 
In particular, we will restrict ourselves to Virasoro Ishibashi states 
associated with such degenerate representations of the Virasoro algebra 
in building up our symmetry-breaking boundary states. 

An irreducible U(1) representation with lowest conformal dimension 
$h_{m,n}=j^2$ for some $j\in\half\Zop$ -- i.e.\  coming from a U(1)$_k$ 
module with label as in \ncondition\ --, decomposes as 
\eqn\decomp{
\H^{U(1)}_{m,n=\nu  \kappa} = \bigoplus_{l=0}^{\infty}\, 
    \H^{\rm Vir}_{|m\kappa + {\nu\over2}|+l} \,,}
where $\H^{\rm Vir}_{p}$ (with $p\geq 0$) denotes the irreducible 
Virasoro representation with highest weight $h=p^2$.
The Virasoro characters for $c=1$ are given by  
\eqn\Vir{\eqalign{
h\ne j^2 \qquad & \chi_h^{\rm Vir}(q) = {q^h \over \eta(q)} \equiv \vartheta_{\sqrt{2h}}(q)\;, \cr
h = j^2  \qquad & \chi_h^{\rm Vir}(q) = \vartheta_{\sqrt{2}j }(q) 
                               - \vartheta_{\sqrt{2}(j+1)}(q)\ .}}
\medskip
Combining left and right movers, the full SU(2)$_k$ state space is given by
\eqn\statespace{\eqalign{\bigoplus_{J=0,\ldots, {k\over 2}}H^{{\rm SU(2)}_k}_J\otimes \bar{H}^{{\rm SU(2)}_k}_J & =
\bigoplus_{{{{{{J=0,\ldots,{k\over 2}\atop{n_L,n_R=-k+1,\ldots, k}}
\atop{2J+n_l{\rm even}}}\atop{2J+n_r{\rm even}}
}}}}H^{{\rm PF}}_{(J,n_L)}\otimes \bar{H}^{{\rm PF}}_{(J,n_R)^+}
\otimes H^{{\rm U(1)}_k}_{n_L}\otimes \bar{H}^{{\rm U(1)}_k}_{{{n_R}}^{+}}\cr
}}
where the subscripts $(J,n)^+ := (J,-n)$ of parafermion and $n^{+} := -n$ of 
U(1)$_k$ representations denote conjugate sectors. 
\hbn
Restricting to  $k = \kappa^2$, we can apply the above decompositions of left- and 
right-moving U(1) representations and obtain an explicit expression for that subspace of 
the WZW bulk space from which we will build up our symmetry-breaking boundary 
states: 
\eqn\newstatespace{\bigoplus_{J=0,\ldots, {k\over 2}}H^{SU(2)_k}_J\otimes \bar{H}^{SU(2)_k}_J\;  \supset\!\!\!\!\!\!\!\!\!\!\!\!
\bigoplus_{{{{{{{{J=0,\ldots, {k\over 2}\atop{\n_L,\n_R=\k+1,\ldots,\k}}\atop{m_L,m_R\in\Zop}}\atop{{\rm s.t.}2J+\n_L{\rm even}}}\atop{{\rm s.t.}2J+\n_R{\rm even}}}}}}\atop{l_L,l_R\in\Zop_+}}\!\!\!\!\!\!\!\!\!\!\!\!
H^{{\rm PF}}_{(J,\n_L\kt)}\otimes \bar{H}^{{\rm PF}}_{(J,-\n_R\kt)}\otimes H^{{\rm Vir}}_{|\,m_L\kappa+{{\n_L}\over 2}|+l_L}\otimes \bar{H}^{{\rm Vir}}_{|\,m_R\kappa-{{\n_R}\over 2}|+l_R}\ .}
We have 
ignored all contributions from U(1)$_k$ modules 
which yield non-degenerate Virasoro representations. Note that this choice also 
influences what PF representations are contained in the subspace given in 
\newstatespace, since the U(1)$_k$ label is coupled to the PF label. 

To be able to form PF$\,\times\,$Vir Ishibashi states, we need 
$(J,\n_L\k) \sim (J,-\n_R\k)$ in the PF part (as Ishibashi states couple 
a representation on the left to its conjugate on the right) 
and $h_L=h_R$ in the Virasoro part, i.e.\ that 
$|m_L\k+{\n\over 2}|+l_L=|m_R\k-{\n\over 2}|+l_R$. 
Switching notations to $r := m_L\k+{\n\over 2}$ and $s:= -m_R\k+{\n\over 2}$, 
we see that the SU(2) bulk space provides a Virasoro Ishibashi state over 
highest weight $j^2$ with $j \in \half\Zop$  whenever $-j \leq r,s \leq j$ and
\eqn\rsvalues{
\quad r+s =\k\rho+\n \ , \quad 
r-s  =\k\rho'\quad\  {\rm for\ some}\ \ \rho,\rho'\in\Zop\quad {\rm with}\ \ \rho+\rho'\ 
{\rm even}.
}
The condition $(J,\n_L\k) \sim (J,-\n_R\k)$ on the parafermion representations 
simply amounts to $\n_R = -\n_L =: \n$, as long as the level $k$ is odd,  or as long 
as $k$ is even and $J \neq {k\over 4}$. 
\sn
For this last case $J = {k\over4}$, however, a complication arises from the 
field identification \fieldid: Since  $({k\over 4}, n_L+k)\sim ({k\over 4}, n_L)$, 
there are additional PF $\,\times\,$Vir Ishibashi states whenever $J= {k\over4}$ 
and $-n_R = n_L+ k$; we call those Ishibashi states $|{k\over4},\nu \kappa\rangle\!\rangle_{\rm tw}$. 

\newsec{Boundary state construction for $r=\sqrt{k}\,r_{{\rm s.d.}}= \k\,r_{{\rm s.d.}}$}

Boundary states can be written as linear combinations of Ishibashi states, 
the latter implementing the gluing conditions of the preserved 
symmetry algebra. For rational models with charge conjugate partition 
function, one can always form the (maximally symmetric) Cardy boundary 
states, where the coefficients in the superposition are given in terms 
of modular $S$-matrix elements. Those boundary states automatically 
satisfy Cardy's conditions, requiring that the overlap of two boundary 
states can be written as an open string partition function \cardy.
\hbn
To obtain symmetry-breaking boundary states, one has to deviate from 
Cardy's construction. We will be guided by the results from 
\refs{\GRW,\GR}\ and \Janik, where boundary states preserving only 
conformal symmetry were presented for $c=1$ models. 
In particular, \GR\ studied free bosons compactified at 
a radius $r={M\over N}\,r_{{\rm s.d.}}$ where $M,\,N$ are coprime integers, 
and found that conformal boundary conditions come in SU(2)$/(\Zop_M\times \Zop_N)$
families
\eqn\bosboundstates{
 |\!|\,g \rra_{c=1} =\ \  2^{-{1\over4}}\sqrt{MN}\!\!\!\sum_{{{j;r,s}\atop{r-s\equiv 0\,({\rm mod}\;M)}}\atop{r+s\equiv 0\,({\rm mod}\;N)}} D^j_{r,s}(g)\; |j;r,s\rra  \ ;  
}
the summation is over $j,r,s\in\half\Zop$ with $j\geq 0$ and $-j\leq r,s\leq j$, and the coefficients \Ham\ 
\eqn\matrixele{
 D^j_{r,s}(g)=\sum_{l=\max(0,s-r)}^{\min(j-r,j+s)}{{[(j+r)!(j-r)!(j+s)!(j-s)!]^{{1\over 2}}} \over (j-r-l)!\, (j+s-l)!\, l!\, (r-s+l)!}\phantom{xxxxx}
}\vskip-7pt  
$$ \hskip130pt \times a^{j+s-l} (a^*)^{j-r-l} b^{r-s+l} (-b^*)^{l}$$
are matrix elements in a spin $j$ representation of SU(2), with $g\in {\rm SU(2})$ 
taken in the form 
\eqn\group{
g= \pmatrix{ a & b \cr - b^\ast & a^\ast} \ .
}
It was shown in \refs{\GRW,\GR} that these boundary states satisfy Cardy's conditions, 
and that for special values of the parameter $g$ they reduce to superpositions 
of Neumann or Dirichlet boundary states. 
\mn
For the SU(2) case we are interested in now, we have obtained, from the SU(2) 
modules, PF$\,\times\,$Vir Ishibashi states, and we can try to combine 
the $c=1$ conformal boundary states from above with Cardy boundary states for 
the parafermionic part. We propose to consider the following boundary 
states for SU(2) WZW models with diagonal bulk partition function: 
\eqn\boundarystate{|\!|\,g_\alpha; J_\alpha,n_\alpha \rra = \ {\cal N}\!
\sum_{{{\nu =-\kappa+1,\ldots,\kappa}\atop{J=0,,\ldots, ,{k\over 2}}}\atop{ 2J+\nu {\kappa\;{\rm even}}}}
\sum_{{{{{j \in{1\over 2}\Zop_+}\atop{r+s=\k\rho+\n}}}\atop{r-s=\k\rho'}}\atop{{\rm s.t.}\ \rho+\rho'{\rm even}}}
B^{{\rm PF}\ (J,\nu \kappa)}_{(J_\alpha,n_\alpha)}\,D^j_{r,s}(g_\alpha)\;
|J,\nu \kappa\rra \otimes |j;r,s\rra
} 
${\cal N}$ is some normalisation factor, the summation range is dictated by the criteria \rsvalues\ ensuring existence of 
degenerate Ishibashi states; furthermore, $g_\alpha \in\,$SU(2) and 
$D^j_{r,s}(g_\alpha)$ are as in \matrixele, while the coefficients 
$B^{{\rm PF}\ (J,\nu \kappa)}_{(J_\alpha,n_\alpha)}$ are as in 
parafermionic Cardy boundary states, i.e.\ 
\eqn\cardycoeff{
B^{{\rm PF}\ (J,\nu \kappa)}_{(J_\alpha,n_\alpha)}=
{S^{{\rm PF}}_ {(J_\alpha,n_\alpha)\,,\,(J,\nu \kappa)}\over 
{\sqrt{S^{{\rm PF}}_{(0,0)\,,\,(J,\nu \kappa)}}}}
}
with the $S$-matrix 
\eqn\smatrixdecomp{S^{{\rm PF}}_{(J,n),(J',n')}=\sqrt{{2\over k}}\,e^{{i\pi nn'\over k}}S^{\suk}_{J,J'}
\quad {\rm with}\quad
S^{\suk}_{J,J'}=\sqrt{{2\over k+2}}\;\sin {(2J+1)(2J'+1)\over k+2}\pi
}
from modular transformations of parafermionic characters (with $q=e^{2\pi i\tau},\ 
\tilde q = e^{-2\pi i/\tau}$) 
$$
\chi^{PF}_{(J,n)}(\tilde q)=
\summandone S^{PF}_{(J,n),(J',n')}\;\chi^{PF}_{(J',n')}({q})\ .
$$
\sn
Our main task in the following is to verify whether the boundary states \boundarystate\ 
satisfy Cardy's condition: We need to compute the overlap of two such boundary states, 
\eqn\overlappp{
\A_{\alpha\beta} =\langle\!\langle \,g_\alpha;J_\alpha,n_\alpha|\!|\, 
 \tilde q^{\half\bigl(L^{SU(2)_k}_0+\bar{L}^{SU(2)_k}_0-{c\over12}\bigr)}\, 
  |\!|\,g_\beta; J_\beta, n_\beta \rangle\!\rangle
}  
and perform a modular transformation to the open string channel. It will turn out 
that the result can indeed be written as a positive integer linear combination 
of parafermion times Virasoro characters, as required. The computation, however, 
is rather lengthy and involves a rather intricate interplay of field identification, symmetries of structure constants and SU(2) group representations. We will 
for simplicity restrict 
to the case that the level $k$ is the square of an odd number $\kappa$ at first. 
The case of even $\kappa$ is discussed at the end of this section, and the 
case where $k$ is not a square (where we will make use of the constructions in \Janik) in section 4. 
\mn
We start our computation of the boundary state overlap $\A_{\alpha\beta}$ by recalling that 
Ishibashi states are orthogonal and normalised in the sense that their 
self-overlap (with the closed string propagator inserted as in \overlappp) 
produces characters of the associated representations. In the present case this 
produces products of parafermion and Virasoro characters, since our Ishibashi 
states have tensor product form. So the overlap is 
$$\A_{\alpha\beta}=\summand\sum_{{{{{j \in{1\over 2}\Zop_+}\atop{r+s=\k\rho+\n}}}\atop{r-s=\k\rho'}}\atop{{\rm s.t.}\ \rho+\rho'\ {\rm even}}} 
\overline{B}{}^{{\rm PF}\ \alpha}_{\bracket}\,B^{{\rm PF}\ \beta}_{\bracket}
\,\chi^{{\rm PF}}_{\bracket}(\tilde{q})\; \virquantity$$
where we have used an abbreviated notation for the parafermion coefficients $B^{\rm PF}$; the bar denotes complex conjugation. 
\hbn
We rewrite this in terms of parafermionic and Virasoro contributions, 
$$
\A_{\alpha\beta}=\summand \A_{\alpha\beta;\; \bracket}^{\rm PF}\ \tilde\A_{\alpha\beta;\; \nu }^{\rm Vir} \ \ ,
$$
with
\eqn\overlapparabit{
\A_{\alpha\beta;\; \bracket}^{\rm PF} = 
\overline{B}{}^{{\rm PF}\ \alpha}_{\bracket}\,B^{{\rm PF}\ \beta}_{\bracket}
\;\chi^{{\rm PF}}_{\bracket}(\tilde{q})
}
and
\eqn\overlapvirbit{
\tilde{\A}_{\alpha\beta;\; \nu }^{\rm Vir} = \sum_{{{{{j \in{1\over 2}\Zop_+}\atop{r+s=\k\rho+\n}}}\atop{r-s=\k\rho'}}\atop{{\rm s.t.}\ \rho+\rho'\ {\rm even}}} \ \virquantity \ \ .
}
We will need to perform the entangled summations over parafermionic and Virasoro 
indices eventually, but first we resolve the constraints on the 
$\rho, \rho'$ summation:  
We note that all the parafermionic constituents in the overlap are invariant 
under the field identification \fieldid, therefore 
$\A_{\alpha\beta;\; \bracket}^{\rm PF} = \A_{\alpha\beta;\; \bracketone}^{\rm PF}$. 
\hbn
On the other hand, the condition $\rho+\rho'$ even in \rsvalues\ changes into 
$\rho+\rho'$ odd under $n \mapsto n+k$, i.e.\ 
$$
\tilde\A_{\alpha\beta;\; \nu  + \kappa}^{\rm Vir} = \sum_{{{{{j \in{1\over 2}\Zop_+}\atop{r+s=\k\rho+\n}}}\atop{r-s=\k\rho'}}\atop{{\rm s.t.}\ \rho+\rho'\ {\rm odd}}} 
\ \virquantity \ \ .
$$
Thus we can rewrite \hphantom{\vbox{\eqn\finaloverlap{}}}
$$\eqalignno{
\A_{\alpha\beta}&= {1\over2} \summand \A_{\alpha\beta;\; \bracket}^{\rm PF}\ 
                            \tilde\A_{\alpha\beta;\; \nu }^{\rm Vir} 
+ {1\over2} \summand \A_{\alpha\beta;\; \bracketone}^{\rm PF}\ 
                           \tilde\A_{\alpha\beta;\; \nu  + \kappa}^{\rm Vir} 
&\cr
&= {1\over2} \summand \A_{\alpha\beta;\; \bracket}^{\rm PF}\ 
  \A_{\alpha\beta;\; \nu }^{\rm Vir}
&\finaloverlap\cr}$$
with $\A_{\alpha\beta;\; \nu }^{\rm Vir}$ defined as in \overlapvirbit\ but 
{\sl without} any restriction on $\rho+\rho'$. 
\sn
To proceed, we first exploit the fact that the Virasoro part is independent of the SU(2) 
spin $J$ in order to perform the $J$ summation in the parafermion contribution -- 
after splitting into integer $J$ (coupled to even $\nkt$ because of the constraint
$2J+\nkt \equiv 0\;({\rm mod}\,2$) and half-odd integer $J\in\Zop+\half$ (coupled 
to odd $\nkt$). We perform a modular transformation of the parafermion 
characters and use the explicit form of the parafermionic coefficients \cardycoeff: 
$$
\sum_{ J \atop {J\in \Zop}} \A_{\alpha\beta;\; \bracket}^{\rm PF}  = 
{2\over k}\;\sum_{ J \atop {J\in \Zop}} 
\summandone e^{ {i\pi\nu \over \kappa}(n'+n_\beta-n_\alpha)}
{{S^{{\rm SU(2)}_k}_{J,J_\alpha} S^{{\rm SU(2)_k}}_{J,J_\beta}S^{{\rm SU(2)_k}}_{J,J'}\over S^{{\rm SU(2)_k}}_{0,J}}\chi^{{\rm PF}}_{(J',n')}(q)}\ .
$$
After inserting $\half\bigl(1+(-1)^{2J}\bigr)$, which projects onto integer $J$,  
and using the identity $(-1)^{2J}S^{{\rm SU(2)}_k}_{J,J'}=
S^{{\rm SU(2)}_k}_{J,{k\over 2}-J'}$, 
the Verlinde formula yields
$$
\sum_{ J \atop {J\in \Zop}} {{S^{{\rm SU(2)_k}}_{J,J_{\alpha}}} 
S^{{\rm SU(2)_k}}_{J,J_{\beta}}S^{{\rm SU(2)_k}}_{J,J'}\over S^{{\rm SU(2)_k}}_{0,J}}
= 
\sum_{ J }  {1+(-1)^{2J}\over 2}\;
 {{S^{{\rm SU(2)_k}}_{J,J_{\alpha}}} S^{{\rm SU(2)_k}}_{J,J_{\beta}}S^{{\rm SU(2)_k}}_{J,J'}\over S^{{\rm SU(2)_k}}_{0,J}}
= \half \Bigl(
N^{J'}_{J_{\alpha},J_{\beta}}+N^{{k\over 2}-J'}_{J_{\alpha},J_{\beta}}\Bigr)\ .
$$
With this, we obtain\hphantom{\vbox{\eqn\pfmodinteger{}}}
$$\eqalignno{
\sum_{ J \atop {J\in \Zop}} \A_{\alpha\beta;\; \bracket}^{\rm PF}  
&=\ {2\over k}\!\!\!\summandone {1\over 2}\biggl(
N^{J'}_{J_{\alpha},J_{\beta}}+N^{{k\over 2}-J'}_{J_{\alpha},J_{\beta}}\biggr)
\;e^{{i\pi\nu \over \kappa}(n'+n_{\beta}-n_{\alpha})}\ 
\chi^{{\rm PF}}_{(J',n')}(q)
&\cr 
&= \ {2\over k}\!\!\!\summandone
e^{{i\pi\nu \over \kappa}(n'+n_{\beta}-n_{\alpha})}\;N^{J'}_{J_{\alpha},J_{\beta}}\ 
\chi^{{\rm PF}}_{(J',n')}(q)\ \ ;
&\pfmodinteger\cr}$$
the simplification in the last step arises after a change of summation variables 
$(J',n') \mapsto ({k\over2}-J',n'+k)$ in the 
$N^{{k\over 2}-J'}_{J_{\alpha},J_{\beta}}$, which leaves the parafermionic characters 
invariant and also the exponential because $\nu $ is even here for integer $J$. 
\hbn
In the same way, one can perform the sum over $J\in\Zop+\half$, 
using the projector
$\half\bigl(1-(-1)^{2J}\bigr)$ and recalling that here $\nu $ is odd; at the 
end of the day, one arrives at the same expression as for integer $J$:  
\hphantom{\vbox{\eqn\pfmodhalfinteger{}}}
$$\eqalignno{
\sum_{ J \atop {J\in \Zop+\half}} \A_{\alpha\beta;\; \bracket}^{\rm PF}  
&=\ {2\over k}\!\!\!\summandone {1\over 2}\biggl(
N^{J'}_{J_{\alpha},J_{\beta}}-N^{{k\over 2}-J'}_{J_{\alpha},J_{\beta}}\biggr)\;
  e^{{i\pi\nu \over \kappa}(n'+n_{\beta}-n_{\alpha})}\ 
\chi^{{\rm PF}}_{(J',n')}(q)
&\cr 
&=\ {2\over k}\!\!\!\summandone
e^{{i\pi\nu \over \kappa}(n'+n_{\beta}-n_{\alpha})}\;N^{J'}_{J_{\alpha},J_{\beta}}\ 
\chi^{{\rm PF}}_{(J',n')}(q)\ \ . 
&\pfmodhalfinteger\cr}$$
\sn
To complete the calculation of the overlap between two symmetry-breaking 
boundary states, we return to the Virasoro contribution to \finaloverlap. We need to evaluate 
\eqn\overlapone{\A_{\alpha\beta;\; \nu }^{\rm Vir} =
  \sum_{{{j;r,s\atop{j\in {1\over 2}\Zop_+}, -j \leq r,s \leq j}
  \atop{r-s\equiv 0({\rm mod}\kappa)}}\atop{r+s\equiv\nu ({\rm mod}\kappa)}}
\overline{D}{}^{j}_{r,s}(g_\alpha)\,D^{j}_{r,s}(g_\beta)\ \chi^{{\rm Vir}}_{j^2}(\tilde{q})
}
for $g_\alpha,g_\beta\in$ SU(2). The restrictions on the permissible $r$ and $s$ can be 
treated in a similar way to \GR\ , leading to the insertion of projection operators 
into the representation matrix $D^j$. To see this, we first implement the mod $\kappa$
requirements on $r,s$ with the help of primitive $\kappa^{\rm th}$ roots of unity: 
\eqn\overlaptwo{\A_{\alpha\beta;\; \nu }^{\rm Vir} =
{1\over{\kt}^2}\sum_{l=0}^{{\kt}-1}\sum_{p=0}^{{\kt}-1}
\sum_{{{j;r,s\atop{j\in {1\over 2}\Zop_+}}\atop{r,s=-j,\ldots, j}}}
e^{{2\pi i\over{\kt}}p(r+s-\nu )}\;e^{{2\pi i\over{\kt}}l(r-s)}
\ \overline{D}{}^{j}_{r,s}(g_\alpha)\,D^{j}_{r,s}(g_\beta)\ \chi^{{\rm Vir}}_{j^2}(\tilde{q}) 
}
Part of the exponentials can be absorbed into the $D^j$ using the matrix 
\eqn\trick{\Gamma_{\kt} = \pmatrix{ e^{{\pi i \over{\kt}}} & 0 \cr 
0 & e^{-{\pi i \over{\kt}}}} \in {\rm SU(2)} 
}
which satisfies 
\eqn\Gammaone{
D^j_{r,s}(\Gamma_{\kt})=e^{{2\pi i\over\kt} r}\delta_{r,s}
}
and hence  
\eqn\Gammatwo{
e^{ {2\pi i\over{\kt}} p(r+s)}  \; D^j_{r,s}(g)
= D^j_{r,s}(\Gamma^p_{\kt}\,g\,\Gamma^p_{\kt})\ , 
\quad
e^{ {2\pi i\over{\kt}} p(r-s)}  \; D^j_{r,s}(g)
= D^j_{r,s}(\Gamma^p_{\kt}\,g\,\Gamma^{-p}_{\kt})\ .
}
Now one can use the representation property of the $D^j$ and perform 
the summation over $r,s$:
\eqn\overlaptwo{
\A_{\alpha\beta;\; \nu }^{\rm Vir} =
{1\over{\kt}^2}\sum_{l=0}^{{\kt}-1}\sum_{p=0}^{{\kt}-1}
\sum_{j\in {1\over 2} \Zop_+} 
e^{-{2\pi i\nu \, p\over{\kappa}}}\  
{\rm Tr}\,D^{j}\Bigl(
    \Gamma^p_{\kt}\, g_\alpha^{-1}\,\Gamma^{p+l}_{\kt}\,g_\beta\,\Gamma^{-l}_{\kt}\Bigr)
\ \chi^{{\rm Vir}}_{j^2}(\tilde{q}) 
}
It remains to perform the modular transformation of the Virasoro characters, 
see e.g.\ \GR\ ,  
and to combine the above expression with \pfmodhalfinteger\ for the parafermionic 
part. The overlap \overlappp\ becomes 
$$\A_{\alpha\beta} = {{\cal N}^2\over 2}\,{1\over{\kt}^2}\;{2\over k}
  \summandone\sum_{{{\nu =-\kappa+1,\ldots,\kappa}\atop{l=0,\ldots, {\kt}-1}}\atop{p=0,\ldots, {\kt}-1}} \sum_{ j\in {1\over 2}\Zop_+} 
  e^{{2\pi i\nu \over 2{\kt}}(n'+n_{\beta}-n_{\alpha}-2p)}N^{J'}_{J_{\alpha},J_{\beta}}\,
\chi^{{\rm PF}}_{(J',n')}(q) $$
\vskip-9pt
$$
\hskip120pt \times \  
{\rm Tr}\,D^{j}\Bigl(
    \Gamma^p_{\kt}\, g_\alpha^{-1}\,\Gamma^{p+l}_{\kt}\,g_\beta\,\Gamma^{-l}_{\kt}\Bigr)
\ \chi^{{\rm Vir}}_{j^2}(\tilde{q})
$$ 
\eqn\overlapallone{
= {\sqrt{2}\,{\cal N}^2\, 2\kappa\over k^2}\summandone\sum_{{{l=0,\ldots, {\kt}-1}\atop{m\in\Zop}}}
N^{J'}_{J_{\alpha},J_{\beta}}\ 
\chi^{{\rm PF}}_{(J',n')}(q)\ \vartheta_{{{-\alpha_{Nl}(g_\alpha,g_\beta)}\over {\sqrt{2}\pi}}+\sqrt{2}m}(q)\ \ .}
\hbn
To obtain the last line in \overlapallone, we have first performed the summation 
over $\nu $, which fixes $p$ to be $N:={n'+n_{\beta}-n_{\alpha}\over 2}$ -- this 
is an integer due to the SU(2) fusion rules and the parafermion condition $2J+n$ even. 
Next, modular transformation of the combination of ${\rm Tr}\,D^j$ and Virasoro 
characters results (see e.g.\ \refs{\MMS,\GR}) in theta functions with highest weight 
given by the angles  $\alpha_{Nl}(g_\alpha,g_\beta)$ defined in terms of the SU(2) trace
\eqn\alphaeqn{
2\cos(\alpha_{Nl}(g_\alpha,g_\beta))=\Tr_{1\over 2}\Bigl( 
 g_\alpha^{-1}\Gamma_{\k}^N\Gamma^l_{\k}\,g_\beta\,\Gamma^{-l}_{\k}\Gamma_{\k}^N\Bigr)\ .
}
Since we are free to choose the overall normalisation factor in the definition \boundarystate\ to be 
$$ 
{\cal N} =  \Bigl({\kappa\over\sqrt{2}}\Bigr)^{3/2}\ ,
$$
the result \overlapallone\ shows that our boundary states for odd $k=\kappa^2$ 
do indeed satisfy 
Cardy's condition, which is the most important and usually most restrictive 
non-linear constraint to be imposed on conformal boundary conditions. 
\bn
We now turn to the case when the {\sl level is an even square}.  
As was pointed out at the end of Section 2, there exist 
additional parafermion Ishibashi states 
$|{k\over4},\nu \kappa\rangle\!\rangle_{\rm tw}$
whenever $k$ is even, namely for $J={k\over 4}$, due to the field identification. 
This implies that the subspace 
\newstatespace\ providing symmetry-breaking Ishibashi states needs be extended 
slightly and becomes 
\eqn\ishibstatespaceone{\bigoplus_{J=0,\ldots,{k\over 2}}H^{SU(2)_k}_J\otimes \bar{H}^{SU(2)_k}_J  \supset
\bigoplus_{{{{{{{J=0,\ldots,{k\over 2}\atop{\n=-\k+1,\ldots,\k}}\atop{{\rm s.t.}2J+\n{\rm even}}}}\atop{m_L,m_R\in\Zop}
}}}\atop{l_L,l_R\in\Zop_+}}H^{{\rm PF}}_{(J,\nkt)}\otimes \bar{H}^{{\rm PF}}_{(J,-\nkt)}\otimes H^{{\rm Vir}}_{|\,\kappa m_L+{{\n}\over 2}|+l_L}\otimes \bar{H}^{{\rm Vir}}_{|\,\kappa m_R-{{\n}\over 2}|+l_R}} 
$$+\bigoplus_{{{{{{\atop{\n=-\k+1,\ldots,\k}}\atop{m_L,m_R\in\Zop}}
}}}\atop{l_L,l_R\in\Zop_+}}H^{{\rm PF}}_{({k\over 4},\nkt)}\otimes \bar{H}^{{\rm PF}}_{({k\over 4},-\nkt+k)}\otimes H^{{\rm Vir}}_{|\,\kappa m_L+{{\n}\over 2}|+l_L}\otimes \bar{H}^{{\rm Vir}}_{|\,\kappa(m_R-{1\over 2})-{{\n}\over 2}|+l_R.}$$
The additional Ishibashi states allow us to refine the boundary states \boundarystate, 
so as to resolve the field identification `fixed point'. To this end, we make the ansatz (cf.\ \MMS)
\eqn\bdystateforkeven{
|\!|\,g,\pm\, \rra_{\rm tot} =
{1\over 2}\,\bigl(\; |\!|\,g\, \rra \pm|\!|\,g\, \rra_{\rm tw} \bigr)\ ,
}
where $|\!|\,g\, \rra $ stands for the boundary state in 
\boundarystate$\,$ and $|\!|\,g\, \rra_{\rm tw}$ is given by 
\eqn\boundarystatetwo{|\!|\,g_\alpha; J_\alpha,n_\alpha\, \rra_{\rm tw} =\ \ {\cal N}_{\rm tw}\!\sum_{{{\nu =-\kappa+1,\ldots,\kappa}}}
\sum_{{{{{j \in{1\over 2}\Zop_+}\atop{r+s=\k\rho+\n+{\k\over 2}}}}\atop{r-s=\k\rho'-{\k\over 2}}}\atop{{\rm s.t. \rho+\rho'{\rm even}}}}
B^{{\rm PF}\ \alpha}_{({k\over4},\nu \kappa)} \;
D^j_{r,s}(g_\alpha)\ 
|{\textstyle {k\over4}},\nu \kappa\rangle\!\rangle_{\rm tw} \otimes |j;r,s\rra
}
with $g_\alpha$ and, for the time being, $(J_\alpha, n_\alpha)$ as in $|\!|\,g_\alpha \rra $. 
Compared to \boundarystate, the possible values of $r$ and $s$ summed over 
in \boundarystatetwo\ have been altered in response to the shift in the 
Virasoro label.
\sn
We need to verify that the overlaps of two boundary states of type 
$|\!|\,g,\pm\,\rra_{\rm tot}$ still satisfy Cardy's conditions. First note that 
$$\langle\!\langle \,g_\alpha|\!|\, q^{\half(L^{SU(2)_k}_0+\bar{L}^{SU(2)_k}_0-{c\over12})}\, |\!|\,g_\beta \rangle\!\rangle_{\rm tw}=0 
$$
since the additional parafermionic Ishibashi states are orthogonal to 
the ones encountered for odd $k$. 
Thus the only new quantity we need to compute is 
$$
\A_{\alpha\beta;\;{\rm tw}} = {}_{\rm tw}\langle\!\langle \,g_\alpha|\!|\, 
 q^{\half(L^{SU(2)_k}_0+\bar{L}^{SU(2)_k}_0-{c\over12})}\,|\!|\,g_\beta \rangle\!\rangle_{\rm tw}\ ,
$$
which can be done along similar lines as for odd $k$, but with 
some differences: The offset by ${\kappa\over2}$ in the $r,s$ 
summation leads to an extra factor $(-1)^{l+N}$ from the 
Virasoro part. In the parafermion contribution, there is no sum 
over $J$, instead $J={k\over4}$ is fixed and one has to 
exploit
$$ 
S^{{\rm SU(2)}_k}_{{k\over4},J} = \cases{ \sqrt{{2\over k+2}}\,(-1)^J &$J$ even\cr \ \ 0 &$J$ odd\cr} \ .
$$
This leads to an additional factor $(-1)^{J_\alpha+J_\beta+J'} = (-1)^N$ in the overlap -- and it also implies that 
$|\!|\,g_\alpha; J_\alpha,n_\alpha\, \rra_{\rm tw}$ is non-zero 
only for integer $J_\alpha$, and that all integer $J_\alpha$ 
lead to the same $|\!|\,g_\alpha; J_\alpha,n_\alpha\, \rra_{\rm tw}$. 
 Altogether, one arrives at 
\eqn\jkoverfour{
\A_{\alpha\beta;\;{\rm tw}} = 
{\cal N}_{\rm tw}^2\;{2\over {k+2}}{1\over{\kt}^2}{2\over k}\sqrt{2}
\sum_{{(J',n')\atop J'\in\Zop,\ n' {\rm even}}} \sum_{{{l=0,\ldots, {\kt}-1}\atop{m\in\Zop}}}
(-1)^{l}\ \chi^{{\rm PF}}_{(J',n')}(q)\,\vartheta_{{{-\alpha_{Nl}(g_\alpha,g_\beta)}\over {\sqrt{2}\pi}}+\sqrt{2}m}(q)
}
where $N ={n'+n_{\beta}-n_{\alpha}\over 2}$ just as in \overlapallone. 
Choosing the normalisation ${\cal N}_{\rm tw}$ such that the prefactor of the sum disappears, we see that the sum of \jkoverfour\ and \overlapallone\ is a sum of characters with positive integer coefficients if and only if we choose $J_\alpha = J_\beta = {k\over4}$ 
in the boundary state $|\!|\,g_\alpha; J_\alpha,n_\alpha\, \rra$; 
to see this, recall that $N^{J'}_{{k\over4},{k\over4}} = 1$ 
precisely if $J'$ is integer. 
\hbn
Summarizing, we find two additional families of boundary states 
for even level $k=\kappa^2$, 
$$
|\!|\,g_\alpha;n_\alpha, n_\alpha^{\rm tw}; \pm\, \rra_{\rm tot} 
={1\over 2}\,
\bigl(\; |\!|\,g_\alpha; {\textstyle {k\over4}}, n_\alpha\, \rra \pm|\!|\,g_\alpha; 0, n_\alpha^{\rm tw}\, \rra_{\rm tw} \bigr)\ .
$$
\bn
\bn
We have thus constructed families of symmetry-breaking 
boundary states for SU(2)$_k$ for any square level $k=\kappa^2$. 
They are parametrised by discrete labels $J_\alpha, n_\alpha$ for
the parafermionic degrees of freedom and SU(2) elements $g_\alpha$
for the Virasoro part. There are, however, identifications within 
those SU(2) families. To see this, note that the representation 
matrix elements $D^j_{r,s}(g)$ from \matrixele\ satisfy
$$
D^j_{r,s}\bigl(\,\Gamma_{\kappa}\,g\,\Gamma_{\kappa}^{-1}\,\bigr)
= e^{{2\pi i\over \kappa}(r-s)}\; D^j_{r,s}(g)\ \ ,
$$
and that they show up in the boundary state only for 
$r-s \equiv 0\;({\rm mod}\, \kappa)$ for $k$ odd and for 
$r-s \equiv {\kappa\over 2}\;({\rm mod}\, \kappa)$ for $k$ even. 
In the former case, the $\Zop_\kappa$-action 
$g \mapsto \Gamma_\kappa g \Gamma_\kappa^{-1}$ leaves 
the boundary state invariant, in the latter case it 
swaps the $|\!|\,g,+\, \rra_{\rm tot}$ with the
$|\!|\,g,-\, \rra_{\rm tot}$ branch.  
This means that instead of SU(2), the paramaters 
$g_\alpha$ in our boundary states \boundarystate\ 
and \bdystateforkeven\ take values in 
$$
g_\alpha \in {\rm SU(2)}/\Zop_{\kappa}\ \ .
$$
Equivalently, we can restrict to the + sign in \bdystateforkeven\
and take $g_\alpha \in {\rm SU(2)}/\Zop_{{\kappa\over 2}}$ for 
even $k=\kappa^2$. 

\newsec{Symmetry-breaking boundary states when $k$ is not a square}

We can also construct boundary states with a reduced PF$\;\times\;$Vir 
symmetry when the SU(2) level $k$ is not a square. Once more starting from 
the decomposition of the SU(2) bulk state space 
\eqn\statespace{\bigoplus_{J=0,\ldots, {k\over 2}}H^{SU(2)_k}_J\otimes \bar{H}^{SU(2)_k}_J=
\bigoplus_{{{{{{{J=0,\ldots, {k\over 2}\atop{n_l,n_r=-k+1,\ldots, k}}\atop{m_l,m_r\in\Zop}}\atop{{\rm s.t.}2J+n_l{\rm even}}}\atop{{\rm s.t.}2J+n_r{\rm even}}
}}}}H^{{\rm PF}}_{(J,n_l)}\otimes \bar{H}^{{\rm PF}}_{(J,n_r)}\otimes H^{U(1)_k}_{n_l}\otimes \bar{H}^{U(1)_k}_{n_r}}
from before, we again concentrate on those U(1) modules which break up into an
infinite number of Virasoro irreducibles as in \decomp. We see from \confdim\ 
that for non-square $k$ this only happens in the vacuum module,  
$h_{(m_l,n_l)}={\bar{h}_{(m_r,n_r)}}=0.$
\hbn
The subspace to which we associate symmetry-breaking Ishibashi states is thus 
\eqn\statespaceone{\bigoplus_{J=0,\ldots, {k\over 2}}H^{SU(2)_k}_J\otimes \bar{H}^{SU(2)_k}_J \supset
\bigoplus_{{{{{J=0,\ldots, {k\over 2}\atop{{\rm s.t.}2J{\rm even}}}\atop{l_L,l_R\in\Zop}}}}}H^{{\rm PF}}_{(J,0)}\otimes \bar{H}^{{\rm PF}}_{(J,0)}\otimes H^{{\rm Vir}}_{l_L}\otimes \bar{H}^{{\rm Vir}}_{l_R}\ ,}
where we have used the explicit decomposition
$H^{U(1)}_{0}=\bigoplus_{l=0}^{\infty}H^{{\rm Vir}}_{l}$ 
to rewrite the chiral U(1) modules in terms of Virasoro modules. 
For these values of $h$ and $\bar{h}$, the SU(2) matrix elements $D^{j}_{r,s}(\,g)$ defined in \matrixele\  become the $j^{\rm th}$ Legendre polynomial $P_j(x)$, 
see \refs{\Janik,\Friedan} and also \GR. This follows from a simple rearrangement of 
\matrixele$\,$ where now $r=s=0$. 
\sn
As a consequence, the boundary states \boundarystate\ have to be altered slightly; we define 
\eqn\boundarystateone{
|\!|\,x; J_\alpha \rra =\ {\cal N}\!\sum_{{{J=0,\ldots, {k\over 2}}}\atop{{\rm s.t.}\, 2J\,{\rm even}}}
\sum_{l\in\Zop_+} \  B^{{\rm PF}\ (J_\alpha,0)}_{(J,0)}\;
P_l(x)\ |J,0\rra \otimes |l\rra \hskip20pt{\rm for}\ x\in[-1,1]\ .
}
Computing the overlap of two such boundary states, and thus verifying that 
they satisfy Cardy's condition, is easier in this case since parafermionic 
and Virasoro parts decouple. Moreover, the field identification \fieldid\ plays 
no role in evaluating the parafermionic contribution since there is no Ishibashi 
state associated to the state space \statespaceone\ with $|J,0\rra=
|{k\over 2}-J,k\rra$.
We find that 
\eqn\irrationalk{\eqalign{\A&=
\langle\!\langle x_\alpha; J_\alpha |\!|\, \tilde 
q^{\half(L_0+\bar L_0-{c\over12})}\,|\!|\,x_\beta;J_\beta \rangle\!\rangle 
\cr
& =\ {\cal N}^2\!\sum_{{{J=0,\ldots, {k\over 2}}}\atop{{\rm s.t.}\, 2J\,{\rm even}}}
\sum_{l\in\Zop_+}
\overline{B}{}^{{\rm PF}\;\alpha}_{(J,0)}\,B^{{\rm PF}\;\beta}_{(J,0)}
\; P_l(x_\alpha)P_l(x_\beta)\;\chi^{{\rm PF}}_{(J,0)}(\tilde q)
\;\chi^{{\rm Vir}}_{l^2}(\tilde q)\cr
& =\ {2{\cal N}^2\over k}\summandone N^{J'}_{J_{\alpha},J_{\beta}}\,
\chi^{{\rm PF}}_{(J',n')}(q)\sum_{l\in\Zop_+}\;
P_l(x_\alpha)P_l(x_\beta)\,\chi^{{\rm Vir}}_{l^2}(\tilde q)\ .\cr}
}
The computation of the Virasoro contribution, 
involving a modular transformation of the Virasoro characters, 
was presented in \Janik:  
\eqn\janikcomp{
\sum_{l\in\Zop_+}P_l(x_\alpha)P_l(x_\beta)\;\chi^{{\rm Vir}}_{l^2}(\tilde q)
={1\over{\sqrt{2}\pi^2}}\int_{0}^{\pi}\!d\phi'\int_{0}^{\pi}\!d\phi\ 
\sum_{n\in\Zop}\vartheta_{{1\over\sqrt{2}}(n+{t\over{2\pi}})}(q)}
where $t$ is defined through (using $x_\alpha =: \cos\theta_\alpha$ etc.)
\eqn\janikcompone{
\cos{t\over 2} =\cos{\theta\over 2}\cos{\phi\over 2}\ , \quad 
\cos\theta  =\cos\theta_\alpha\cos\theta_\beta-\sin\theta_\alpha\sin\theta_\beta\cos\phi'\ .
}
The final result for the overlap between two symmetry-breaking 
boundary states $|\!|x_\alpha;J_\alpha\rra$ is therefore a 
continuous band spectrum 
\eqn\irrationalkfinal{\A ={2{\cal N}^2\over {k\sqrt{2}\pi^2}}\int_{0}^{\pi}d\phi_2\int_{0}^{\pi}d\phi\summandone\sum_{n\in\Zop} N^{J'}_{J_{\alpha},J_{\beta}}\;
\chi^{{\rm PF}}_{(J',n')}(\tilde{q})\,\chi^{{\rm Vir}}_{{1\over 4}(n+{t\over{2\pi}})^2}(\tilde{q})\ . }

\newsec{Comparison with maximally symmetric SU(2)$_k$ boundary states}

In this section, we would like to investigate the relation between 
our symmetry-breaking boundary states and others constructed previously 
in the literature, in particular maximally symmetric 
boundary conditions for the SU(2) WZW model. The latter are given 
by ordinary Cardy states and by rotated Cardy branes, which can 
be written as (see e.g.\ \RSmod)
\eqn\rotbdystate{ 
|\!|\,J_\alpha \rra_{\lambda_a} = 
\exp\{i\, \lambda_a\,J^a_0\}\;|\!|\, J_\alpha \rra_0
}
where $|\!|\,J_\alpha \rra_0$ is a Cardy boundary state and 
$\lambda_a$, $a=1,2,3$, are rotation parameters. These states
preserve a full SU(2)$_k$ symmetry (albeit a twisted one) 
due to the rotated gluing conditions 
$$
\bigl({\rm Ad}_h (J^b_m) + \bar J^b_{-m}\bigr)\;
|\!|\,J_\alpha \rra_{\lambda_a} = 0
$$
involving the adjoint action of the group element 
$h:= \exp\{\lambda_a t^a\}$ on the currents $J^b_m$. 
Semi-classically, these branes correspond to conjugacy 
classes in the WZW target \AlSchom. 
\hbn
Since the affine Lie algebra preserved by 
$|\!|\,J_\alpha \rra_{\lambda_a}$ is rotated relative to the 
one preserved by Cardy boundary states, the overlap of two branes 
with non-zero relative angle no longer decomposes into full 
affine Lie algebra characters, but is close to the overlaps 
we found in Section 3 for level $k = \kappa^2$. (We will restrict 
to square level in the following and assume that $k$ is odd for 
convenience.)
\sn
We will now show that, for the choice 
$g_\alpha = {\bf 1}_2$ and $n_\alpha=0$, our boundary states 
$|\!|\,g_\alpha; J_\alpha,n_\alpha \rra$ from eq.\ \boundarystate\ 
coincide with superpositions of rotated Cardy branes (with 
Cardy label $J_\alpha$). 
Let us specialise to $\lambda_a = \lambda\,\delta_{a,3}$ in 
\rotbdystate\ and consider the action of the rotation on 
the SU(2)$_k$ Ishibashi states, which can be decomposed into 
parafermion and U(1) Ishibashi states: 
\eqn\rotIshi{\eqalign{
\exp\{i\,\lambda J^3_0\}\,|J\rra^{\rm SU(2)} 
&= 
\sum_{n,m; 2J+n\;{\rm even}}\ |J,n\rra^{\rm PF}
   \otimes \;\exp\{i\,\lambda J^3_0\}\, |n,m\rra^{\rm U(1)}
\cr
&= \sum_{n,m; 2J+n\;{\rm even}}\ |J,n\rra^{\rm PF}
   \otimes \;e^{i\,\lambda\; q_{n,m}}\, |n,m\rra^{\rm U(1)}
\cr}}
(Note that no phases or unitary transformations occur in the 
decomposition of $|J\rra$ into PF$\,\times\,$Vir Ishibashi 
states, since an Ishibashi state $|i\rra$ can be regarded 
as a projector in End${\cal H}_i)$, as follows from \Ishi\ 
and the arguments presented in \refs{\Behthree\PetkovaZuber}.) 
We have used that the U(1) current $J^3_0$ 
commutes with the parafermion sector and is diagonal on each 
U(1) irrep, producing the charge $q_{n,m} = 2\,k\;(m+{n\over 2k})$. 
\hbn
Now let us sum over a discrete set of rotation angles 
$\lambda_l = {2\pi\,l\over \kappa}$ for $l=0,1,\ldots,\kappa-1$. 
Due to the U(1) charges present in the SU(2)$_k$ theory, the 
$l$-summation projects out those U(1) Ishibashi states that 
do not satisfy $n=\nu\,\kappa$ and leaves only those that 
yield degenerate Virasoro representations -- which were 
precisely the ones we restricted to in our ansatz for 
symmetry-breaking boundary states \boundarystate. 
Furthermore, it is straightforward to see that the superposition of 
$\kappa$ rotated Cardy branes $|\!|\,J_\alpha\rra_{\lambda_l}$ 
coincides with the symmetry-breaking boundary state 
$|\!|\,g_\alpha={\bf 1}_2; J_\alpha,n_\alpha=0 \rra$
up to overall normalisation. That the latter agrees, as well, 
can be seen by counting the number of identity operators in 
the partition function of the superposition and in self-overlap 
\overlapallone\  computed in Section 3. \footnote{*}{Note that the 
decomposition \rotIshi\ also allows to show that the overlap 
between $|\!|\,g_N; J_\alpha,n_\alpha \rra $ and an SU(2)$_k$ 
Cardy boundary state yields a good partition function. One needs 
to observe that, from the latter, only those PF$\,\times\,$U(1) 
Ishibashi states which have $\lambda_l\,q_{m,n}\in 2\pi\Zop$ 
contribute to the overlap.}
\sn
For generic $g_\alpha$, the boundary states \boundarystate\ 
cannot be written as superpositions of rotated Cardy branes, 
but for another special choice of the $g_\alpha$ parameters, 
namely
$$
g_\alpha = g_N := \pmatrix{\;0 &1\cr-1&0\cr}\ ,
$$
they resemble the `B-type' boundary states constructed 
in \MMS, see also \refs{\MMStwo,\QS}: One can easily show that 
\eqn\oppositeendbdystate{
|\!|\,g_N; J_\alpha,n_\alpha \rra  \sim
\sum_{J;\ 2J\;{\rm even}} B^{{\rm PF}\ (J, 0)}_{(J_\alpha,n_\alpha)}
\  |J,0\rra^{\rm PF} \otimes |\!| N\rra
}
where the last factor is a superposition of $\kappa$ free boson 
Neumann boundary states with evenly spaced Wilson lines, see 
\refs{\GRW,\GR}. In contrast to the B-type states from \MMS, the 
boundary states \oppositeendbdystate\ only respect PF$\,\times\,$U(1)
symmetry, not PF$\,\times\,$U(1)$_k$. However, the arguments 
from \MMS\ supporting the semi-classical interpretation of 
B-type branes as three-dimensional objects seem applicable to 
\oppositeendbdystate, as well. 
\hbn
Taking this for granted, we are led to conclude that our 
families of symmetry-breaking boundary states interpolate 
between superpositions of (generically 2-dimensional) rotated 
Cardy branes (for $g_\alpha = {\bf 1}_2$) and ``3-dimensional'' 
branes for $g_\alpha=g_N$. At 
both these points, a PF$\,\times\,$U(1) symmetry is preserved, 
while the symmetry is broken down to PF$\,\times\,$Vir for 
generic $g_\alpha$, and no-where enhanced to full SU(2)$_k$.


%
\sn
Note that the `Legendre boundary states' \boundarystateone\ for 
non-integer $\sqrt{k}$ show a similar behaviour at the endpoints 
$x=\pm1$ of the continous parameter: For $x=1$, the Virasoro 
contributation can be written as an integral over ordinary 
Dirichlet boundary states, while $x=-1$ corresponds to an 
integral of Neumann boundary states over the dual circle, see 
\Janik. 
\def\SU{{\rm SU}}
\newsec{Extension to other group targets}

\noindent Building on the ideas of \MMS\ and \QS, it is rather 
straightforward to generate symmetry-breaking boundary states 
for higher rank WZW models from the ones for SU(2) presented 
above. SU(2) can be embedded into any compact Lie group, and 
we can use cosets of the form 
$$G=G/SU(2)\ \times\; SU(2)$$
to break the underlying $G_k$ symmetry. One can, e.g., build 
boundary states of the type
\eqn\generalcoset{|\!|\,(\rho_\a,J_\a) \rra =\sum_{(\mu,J)}B^{(\rho_\a,J_\a)}_{(\mu,J)}
\ |\mu,J\rra \otimes |J\rra}
where $|\mu,J\rra$ and $|J\rra$ in \generalcoset\ are Ishibashi states 
of the coset $G/SU(2)_k$ and $SU(2)_k$ theories respectively; the 
possible $(\mu,J)$ in the summation may be subject to non-trivial field identification and branching selection rules, as seen before in the PF theory. 
The coefficients $B^{(\rho_\a,J_\a)}_{(\mu,J)}$ have to be chosen such 
that Cardy's condition is satisfied, and one possibility 
was presented in \QS, namely 
\eqn\quellaschomerus{
B^{(\rho_{\a},J_\a)}_{(\mu,J)}={S^G_{\rho_\a\mu}\over {\sqrt{S^G_{0\mu}}}}
\;{\overline{S}{}^{{\rm SU(2)}_k}_{J_\a J}\over {\overline{S}{}^{{\rm SU(2)}_k}_{0 J}}}\ .}
This uses the modular $S$-matrices of the $G$ and SU(2) WZW theories, 
and in general leads to a spectrum without $G$-symmetry even though 
trivial gluing conditions apply in the coset and in the SU(2) factor. 
\sn
We can reduce the symmetry even further by incorporating our boundary 
states \boundarystate\ into the decomposition \generalcoset, in place 
of the SU(2) Ishibashi states. We simply modify \generalcoset\ by
decomposing the latter Ishibashi states into ones for PF$\,\times\,$Vir 
as before and propose symmetry-breaking boundary states for the 
$G$ WZW model of the form 
\eqn\coset{
|\!|\,\rho_\a, J_\a, n_\a g_a \rra = {\cal N}\;
\sum_{(\mu,J)}\;\summandthree\virsum 
B^{\alpha}_{(\mu,J,\nkt;j,r,s)}\ 
|\mu,J\rra\otimes|J,\nkt\rra\otimes|j;r,s\rra\ ; }
here we assume that the level $k=\kappa^2$ is a square, and 
the coefficients $B^{\alpha}_{(\mu,J,\nkt;j,r,s)}$ are given by 
\eqn\cosetcoeff{B^{\a}_{(\mu,J,\nkt;j,r,s)}=
{S^G_{\rho_\a\mu}\over {\sqrt{S^G_{0\mu}}}}\;
{\overline{S}^{PF}_{(J_\a,n_\a),(J,\nkt)}\over \overline{S}^{PF}_{(0,0),(J,\nkt)}}\;D^j_{r,s}(g)\ .
}
For convenience, we assume that the permissible representations of 
the $G/{\rm SU(2)}$ coset theory are not subject to any field identifications and hence there are no branching selection rules, unlike the PF theory. Although this might restrict the possible coset models at our disposal, there are still many models (e.g.\ the SU(3)/SU(2) theory)
for which this assumption holds and hence for which our construction will 
still be valid. This assumption will simplify the resulting calculation 
since we can split the sum over $(\mu, J)$ in \coset\ into two sums; one 
running over the possible irreducible representations of $G$, the other 
over those of ${\rm SU(2)_k}$. The remaining constraint 
$2J+\nkt \equiv 0\;({\rm mod}\,2)$ can be dealt with as before
-- which also suggests that 
our simplifying assumption on the $G/\SU$ coset can be relaxed. 
\sn
In essence, the computations required to verify Cardy's constraint 
proceed very much along the lines of Section 3. The only additional 
pieces of information needed to calculate the overlap are:

\noindent (1) The $S$-matrix $S^{G/{\rm SU(2)}_k}_{(\mu,J),(\mu'',J'')}$
for modular transformation of coset characters 
may be decomposed into the S matrices of the two independent theories through the formula \break\noindent (see e.g.\ \refs{\FuchsLect,\QS}) 
\eqn\gsu{S^{G/{\rm SU(2)}_k}_{(\mu,J),(\mu'',J'')}=
S^G_{\mu,\mu''}\,\overline{S}^{{\rm SU(2)_k}}_{J,J''}\ .}
\vskip8pt

\noindent (2) The ratios $S^{{\rm SU(2)_k}}_{J_\a,J}/\ssu_{0,J}$  
of $S$-matrix elements (the generalised quantum dimensions) 
form a representation of the fusion algebra, i.e.\
\eqn\fusionalg{
{\ssu_{J_\a,J}\over {\ssu_{0,J}}}\; {\ssu_{J_\b,J}\over\ssu_{0,J}}
=\sum_{J^{''}}N^{J^{''}}_{J_\a,J_\b}\;{\ssu_{J^{''},J}\over \ssu_{0,J}}.}
\vskip8pt

\noindent The overlap of two boundary states as in 
\coset, properly normalised, is then given by
\setbox13\vbox{{\eqn\cosetoverlap{...}}}
$$\eqalignno{
\langle\!\langle \,\a&|\!|\, q^{\half(L_0+\bar{L}_0-{c\over12})}\, |\!|\, 
\b \rangle\!\rangle
&\cosetoverlap\cr
&=\!\! 
\sum_{ 
{ (\mu,J),\; J'',\; (J',n') 
\atop {l=0,\ldots, \kt-1},\ m\in\Zop} 
}
\!\! N^{G\;\mu}_{\rho_{\alpha},\rho_{\beta}}N^{J''}_{J_{\alpha},J_{\beta}}
N^{J}_{J',J''}\;
\chi^{G/\SU(2)_k}_{(\mu, J)}(q)\,\chi^{{\rm PF}}_{(J',n')}(q)\,
\vartheta_{{{-\alpha_{Nl}(g_\a,g_\b)}\over {\sqrt{2}\pi}}+\sqrt{2}m}(q)\ .
&\cr}$$
\mn
It is easy to generalize this construction to produce other
symmetry-breaking boundary states for the Lie group $\SU(M)$ by 
realising it as a string of cosets of the 
form 
\eqn\sundecomposition 
{\eqalign{
\SU(M)& \cong \bigl(\SU(M)/\SU(M-1)\bigr)\times \bigl(\SU(M-1)/\SU(M-2)\bigr)\times\phantom{xxxxxxx} \cr
&\hskip120pt \times\cdots\times \bigl(\SU(3)/\SU(2)\bigr)\times \SU(2)\ . 
\cr}}
Again, one has to be aware of the possibililty of field identification and branching selection rules appearing in the corresponding decomposition of the WZW state space; we can avoid this problem, however, if we demand that the generators of $\SU(l-1)$ form the upper-left block in $\SU(l)$. 
\hbn
The sequence of embeddings can be exploited to produce symmetry-breaking boundary states, as discussed in \QuellaNS. We can again generalise 
the construction there by using one of our symmetry-breaking 
boundary states for the last SU(2) factor. We arrive at 
\def\sumfive{\sum}
\eqn\gencosetboundarystate{
|\!|\alpha\rra=
\sumfive_{{(\rho_M,\rho_{M-1}),\ (\rho_{M-1},\rho_{M-2}), \cdots} \atop (J,\nkt),\ j;r,s}   \!\!
B^{\a}_{(\rho_{ M},\cdots,j;r,s)}\ 
|\rho_M,\rho_{M-1}\rra \otimes 
\cdots \otimes |J,\nkt \rra \otimes |j;r,s\rra 
}
where $(\rho_l,\rho_{l-1})$ denote the irreducible representations of 
the coset $\SU(l)/\SU(l-1)$, and where the coefficients are
\eqn\gencosetcoeff{
B=
\frac{S^{\SU(M)}_{(\rho_M)_{_{\a}}\rho_M}}{{\sqrt{S^{\SU(M)}_{0,\rho}}}}
\,\frac{\overline{S}^{\SU(M-1)}_{(\rho_{M-1})_{\a},\rho_{M-1}}}{\overline{S}^{\SU(M-1)}_{0,\rho_{M-1}}}\,
\frac{\overline{S}^{\SU(M-2)}_{(\rho_{M-2})_{\a},\rho_{M-2}}}{\overline{S}^{\SU(M-2)}_{0,\rho_{M-2}}}\, \cdots \, 
\frac{\overline{S}^{\rm PF}_{(J,\nkt),(J_\a,n_\a)}}{\overline{S}^{\rm PF}_{(0,0),(J,\nkt)}}\;
D^j_{r,s}(g)\ .
}
\noindent 
Computation of the overlap of two such boundary states leads to 
$$\eqalign{  
Z_{\a \b}(q)&=\!\!\sum_{
 { {  (\rho'_M,\rho'_{M-1}),\  (\rho''_{M-1},\rho'_{M-2}) \cdots
            \atop (J'', n')}
                           \atop l=0,\cdots, \kt-1;\ m\in\Zop}
                           } 
   \sum_{
    {{\rho'''_{M-1}, \ \rho'''_{M-2} \cdots} 
                         \atop J'''}   }
N^{\rho'_M}_{ {(\rho_M)}_\alpha {(\rho_M)}_\beta } 
\prod_{l=2}^{M-1}
N^{\rho'''_{l}}_{{(\rho_l)}_\alpha {(\rho_l)}_\beta}
N^{\rho'''_l}_{{\rho'_l} {\rho''_l}}
\cr
&\hskip30pt\times\vphantom{\sum_0^{L^l}}\chi^\frac{M}{{M-1}}_{(\rho'_{M},\rho'_{M-1})}(q)
\chi^{\frac{M-1}{{M-2}}}_{(\rho''_{M-1},\rho'_{M-2})}(q)
\cdots
\chi^{\rm PF}_{(J'',n')}(q)\;\vartheta_{\frac{{-\alpha_{Nl}(g_\a,g_\b)}}{{\sqrt{2}\pi}}+\sqrt{2}m}(q)\cr}
$$
where $\chi^{\frac{M-1}{{M-2}}}_{(\mu'',\nu')}(\tilde{q})$ are 
characters of the coset $\SU(M-1)/\SU(M-2)$.
\hbn

\newsec{Conclusions and open questions}

We have presented a construction of boundary states for 
SU(2) WZW models that break the symmetry to PF$\,\times\,$Vir, 
using a coset decomposition of SU(2) and the most general 
conformal boundary states for a free boson. The moduli 
space of those boundary states depends on whether the 
level $k$ of the SU(2) theory is a square or not: in the 
latter case, we find families parametrised by $x\in [-1,1]$, 
for $k=\kappa^2$ we find a discrete quotient of SU(2), namely 
SU(2)$/\Zop_{\kappa\over2}$ for even $k$ and SU(2)$/\Zop_\kappa$
for odd $k$. At special points of those 3-dimensional families, 
our boundary states are superpositions of $\kappa$ Cardy 
branes at relative angles, for generic parameters they are 
elementary branes -- and so far lack a target space interpretation
(which might be identified using the methods of {\refs{\FFFS,\MMS}).  
\mn
We have performed the most important consistency check and 
shown that the new symmetry-breaking boundary states satisfy 
Cardy's condition, but one obvious open question is to 
check other sewing relations like the cluster condition. 
For the Legendre boundary states from Section 4, this follows 
directly from the analysis given in \Janik, but in the case 
of $k=\kappa^2$ one needs to deal with non-trivial 
field identification issues. 
\mn
It would be interesting to see whether the three-dimensional 
family parametrised by $g_\alpha$ in \boundarystate\ is one of  
marginal boundary deformations. Already the counting of 
boundary operators with conformal dimension one that are 
supported by these boundary states is involved, since the 
spectrum displays a complicated $g_\alpha$-dependence due to
\alphaeqn. However, one can check that there are always at 
least three marginal fields in the spectrum \overlapallone: 
the parafermion vacuum (labelled $J'=n'=0$) tensored with 
dimension one states counted by the theta functions $\vartheta_0(q)$
and $\vartheta_{\pm\sqrt{2}}(q)$. Whether the associated operators 
are truly marginal and responsible for the three-dimensional 
family of symmetry-breaking boundary states is, of course, 
much more difficult to decide. 
\mn
There are somewhat related questions concerning renormalisation 
group flows: All maximally symmetric branes in WZW models (Cardy 
branes as well as rotated ones) can be viewed as condensates of 
D0-branes -- in particular they arise as (perturbative) 
RG fixed points, or as solutions to the equations of motion of 
the effective action computed in \ARS, see also 
\refs{\FredSchom,\Fredenhagen}. It is known, at least for 
higher rank groups, that also symmetry-breaking branes can arise 
from such condensation processes \Monnier -- in fact, the ground 
state of the effective action does not preserve the maximal 
symmetry -- but it is unclear whether the boundary states 
constructed in this paper have a (perturbative) description 
for large $k$. 
\eject\noindent 
On the other hand, since their construction starts from a 
brute force breaking of the symmetry to PF$\,\times\,$U(1), 
it should be relatively straightforward to study their behaviour 
under marginal bulk deformation of the SU(2)$_k$ theory under 
$J^3(z)J^3(\bar z)$, see e.g.\ \refs{\HassanGI,\ForsteKM}. 
In view of the recent results of \FredGab\ on bulk deformations 
of conformal U(1) boundary states, one would not expect our 
symmetry-breaking SU(2) boundary states to decay into other 
boundary conditions under such marginal bulk deformations. 
\mn
While we have given simple generalisations to higher rank
group targets exploiting embeddings of SU(2), one may 
conjecture that there exist other boundary states 
constructed via a decomposition of the $G$ symmetry in which 
the Virasoro algebra from the SU(2) case is extended to a 
higher Casimir W-algebra, e.g.\ $W(2,3,\ldots,N)$ for SU(N). 
It is also tempting to suggest that for higher rank groups $G$, 
one can construct symmetry-breaking boundary states which 
involve representation matrix elements of $G$ in place of 
the $D^j_{r,s}(g)$ for $g\in\,$SU(2). As a first step to 
verify this, one should analyse the $G$ WZW model for level 
$k=1$ in detail. 
\mn
What may be more relevant for applications to string theory 
is a generalisation to supersymmetric WZW models on the one 
hand (in particular to supersymmetric SU(2), which shows up 
in the world-sheet description of NS 5-branes \refs{\CHS,\BianchiGT}), 
and to coset models on the other, which would open up the 
possibility of constructing new families of boundary states 
in CFTs which are important for string model building.

\bn
{\bf Acknowledgments: }We are indebted to Thomas Quella for 
numerous comments and invaluable discussions. We also thank 
S.\ Fredenhagen, M.\ Gaberdiel, A.\ Konechny, V.\ Schomerus 
and G.\ Watts for their interest and useful remarks. 
This work was supported by PPARC under grant numbers PP/C5071745/1
and PPA/S/S/2003/03643, the
EU "European Superstring Theory Network" grant MRTN-CT-2004-512194 and the 
EU network "EUCLID" grant HPRN-CT-2002-00325.

\vfil\eject

\footatend
\immediate\closeout\rfile\writestoppt
\baselineskip=14pt\centerline{{\bf References}}\bigskip{\frenchspacing%
\parindent=20pt\escapechar=` \input refs.tmp\vfill\eject}\nonfrenchspacing

\bye